\documentclass[12pt,a4paper,preprint,superscriptaddress]{revtex4-1}

\usepackage[caption=false]{subfig}
\usepackage{float}
\usepackage{amsmath,amssymb}
\usepackage[utf8]{inputenc}
\usepackage{graphicx}

\usepackage[english]{babel}
\makeatletter\AtBeginDocument{\let\@elt\relax}\makeatother
\makeatletter
\def\bbl@set@language#1{%
  \edef\languagename{%
    \ifnum\escapechar=\expandafter`\string#1\@empty
    \else\string#1\@empty\fi}%
  \@ifundefined{babel@language@alias@\languagename}{}{%
    \edef\languagename{\@nameuse{babel@language@alias@\languagename}}%
  }%
  \select@language{\languagename}%
  \expandafter\ifx\csname date\languagename\endcsname\relax\else
    \if@filesw
      \protected@write\@auxout{}{\string\select@language{\languagename}}%
      \bbl@for\bbl@tempa\BabelContentsFiles{%
        \addtocontents{\bbl@tempa}{\xstring\select@language{\languagename}}}%
      \bbl@usehooks{write}{}%
    \fi
  \fi}
\newcommand{\DeclareLanguageAlias}[2]{%
  \global\@namedef{babel@language@alias@#1}{#2}%
}
\makeatother

\DeclareLanguageAlias{en}{english}

\usepackage{xcolor}

\usepackage{hyperref}

\begin{document}

\title{An efficient plasma-surface interaction surrogate model for sputtering processes based on autoencoder neural networks}
\date{\today}
\author{Tobias Gergs}
\affiliation{Chair of Applied Electrodynamics and Plasma Technology, Department of Electrical Engineering and Information Science, Ruhr University Bochum, 44801 Bochum, Germany}
\affiliation{Electrodynamics and Physical Electronics Group, Brandenburg University of Technology Cottbus-Senftenberg, Siemens-Halske-Ring 14, 03046 Cottbus, Germany}
\author{Borislav Borislavov}
\affiliation{Electrodynamics and Physical Electronics Group, Brandenburg University of Technology Cottbus-Senftenberg, Siemens-Halske-Ring 14, 03046 Cottbus, Germany}
\affiliation{Chair of Communications Engineering, Brandenburg University of Technology Cottbus-Senftenberg, Siemens-Halske-Ring 14, 03046 Cottbus, Germany}
\author{Jan Trieschmann}
\affiliation{Electrodynamics and Physical Electronics Group, Brandenburg University of Technology Cottbus-Senftenberg, Siemens-Halske-Ring 14, 03046 Cottbus, Germany}

\begin{abstract}
Simulations of thin film sputter deposition require the separation of the plasma and material transport in the gas-phase from the growth/sputtering processes at the bounding surfaces (e.g., substrate and target). Interface models based on analytic expressions or look-up tables inherently restrict this complex interaction to a bare minimum. A machine learning model has recently been shown to overcome this remedy for Ar ions bombarding a Ti-Al composite target \cite{kruger_machine_2019}. However, the chosen network structure (i.e., a multilayer perceptron, MLP) provides approximately 4 million degrees of freedom, which bears the risk of overfitting the relevant dynamics and complicating the model to an unreliable extend. This work proposes a conceptually more sophisticated but parameterwise simplified regression artificial neural network for an extended scenario, considering a variable instead of a single fixed Ti-Al stoichiometry. A convolutional $\beta$-variational autoencoder is trained to reduce the high-dimensional energy-angular distribution of sputtered particles to a latent space representation of only two components. In addition to a primary decoder which is trained to reconstruct the input energy-angular distribution, a secondary decoder is employed to reconstruct the mean energy of incident Ar ions as well as the present Ti-Al composition. The mutual latent space is hence conditioned on these quantities. The trained primary decoder of the variational autoencoder network is subsequently transferred to a regression network, for which only the mapping to the particular latent space has to be learned. While obtaining a competitive performance, the number of degrees of freedom is drastically reduced to 15,111 (0.378~\% of the MLP) and 486 (0.012~\% of the MLP) parameters for the primary decoder and the remaining regression network, respectively. The underlying methodology is very general and can easily be extended to more complex physical descriptions (e.g., taking into account dynamical surface properties) with a minimal amount of data required.
\end{abstract}

\maketitle

\newpage

%%%%%%%%%%%%%%%%%%%%%%%%%%%%%%%%%%%%%%%%%%%%%%%%%%%
\section{Introduction}
\label{sec:introduction}

Thin film deposition by sputtering is driven by the interplay of two different states of matter (i.e., solid and plasma). The length and time scales of the intrinsic processes span several orders of magnitude, as detailed elsewhere \cite{kruger_machine_2019, bird_molecular_1994, lieberman_principles_2005, callister_materials_2013}. Consequently, both subsystems and their interaction with one another are commonly studied in terms of separate modeling and computer simulation approaches. Concerning the process of sputtering (i.e., the projectile bombardment of a target surface that leads to a collision cascade in the solid and eventually to the release of atoms), transport of ions in matter (TRIM), Monte Carlo (MC) or molecular dynamics (MD) simulations are typically used \cite{biersack_monte_1980, eckstein_sputtering_1984, moller_tridyn_1984, voter_introduction_2007, graves_molecular_2009, neyts_molecular_2017}. The collisional transport through the plasma (including ionization processes) as well as the plasma itself are usually described by fluid models or particle in cell/Monte Carlo collision (PIC/MCC) simulations, coupled to test-particle method (TPM) or direct simulation Monte Carlo (DSMC) simulations, respectively \cite{birdsall_plasma_1991, dijk_plasma_2009, serikov_particle--cell_1999, somekh_thermalization_1984, turner_monte_1989, trieschmann_transport_2015}. The appropriate models need to be selected based on gas pressure regime as well as physical fidelity of the models (e.g., kinetic vs continuum description). The deposition of sputtered particles is then again either modeled by molecular dynamics, Monte Carlo or hybrid simulations \cite{voter_introduction_2007, graves_molecular_2009, neyts_molecular_2017, neyts_catalyzed_2010, neyts_combining_2012, tonneau_tioxdeposited_2018}.

Joint simulation frameworks that pursue to consistently simulate the plasma and the surface dynamics, substitute detailed surface models by surrogate models for instance based on analytical expressions (e.g., Sigmund-Thompson theory \cite{thompson_ii_1968, sigmund_theory_1969, sigmund_theory_1969-1}), concentrated coefficients (e.g., sputter yield), or interpolation from look-up tables. However, these approaches inherently require drastic simplifications of the complex interactions (in particular in case of reactive plasmas) \cite{tonneau_tioxdeposited_2018}. In addition, establishing an interface model by manually prioritizing and implementing all relevant interactions becomes a tedious task \cite{berg_fundamental_2005, depla_reactive_2008}. In contrast, machine learning models (e.g., artificial neural networks, ANNs), have been used to generalize complex correlations and create surrogate models in the frame of plasma or gas-phase interactions with surfaces \cite{diaw_multiscale_2020, ulissi_address_2017, kino_characterization_2021}. In particular, the feasibility and accuracy of this concept related to the prediction of energy-angular distributions (EADs) of sputtered particles as a function of the impinging projectile ion energy distribution (IED) has been demonstrated \cite{kruger_machine_2019}. The required data has been gathered for Ar ion bombardment on a Ti-Al composite surface (fixed stoichiometry $x=0.5$), which has been simulated with TRIDYN \cite{moller_tridyn_1984}. To mimic scenarios with more complex processes or material systems, which require more fundamental physics models (e.g., MD) and consequently bring about a higher computational load, an intentionally low statistical quality and a small data set size of 439 samples has been chosen. 
% Hence, the training data are comprised of statistically fluctuating representations. An estimate of the ground truth has been made by also evaluating a single simulation with $10^6$ projectiles. This allowed to evaluate the networks' generalization capability. 
A multilayer perceptron (MLP) network type has been chosen due to its methodical simplicity. Despite the ANN's capability to generalize well, due to its large number of degrees of freedom (i.e., approximately 4 million), the model may rather be considered suboptimal with respect to the inherent risk of overfitting. This holds in particular for the targeted scenario of even more challenging data set sizes and statistical representations.\cite{kruger_machine_2019}

This work sets out to improve on the beforehand outlined concept in a threefold way, namely by i) consideration of an extended data set with Ti-Al stoichiometry as an additional surface state parameter, ii) dimensionality reduction by means of convolutional $\beta$-variational autoencoders \cite{kingma_auto-encoding_2013, rezende_stochastic_2014, higgins_beta-vae_2016, burgess_understanding_2017, doersch_tutorial_2021} conditioned on given input parameters, and iii) transfer learning of the pretrained decoder network which exploits the established reduced parameter space. The manuscript is structured as follows: In Section~\ref{sec:methods}, applied methods and parameters are described. The data set generation which encompasses the sputtering simulations as well as their distribution among subsets required for the machine learning approach are introduced in Section~\ref{ssec:data}. The dimensionality reduction and regression models are described in Sections~\ref{ssec:Dimensionality_reduction} and \ref{ssec:Regression_model}, respectively. The corresponding results are gathered in Section~\ref{sec:results}. Finally, the work is summarized and conclusions are drawn in Section~\ref{sec:conclusion}.

\section{Methods}
\label{sec:methods}

In the following, it is first explained how the considered data set is established, which comprises input-output relations (i.e., projectile IEDs to sputtered particle EADs) from physical sputtering simulations using TRIDYN simulations and how it is split into respective subsets for the machine learning procedure. Second, a dimensionality reduction of the species-dependent EADs is described based on an asymmetric convolutional $\beta$-variational autoencoder ($\beta$-VAE). Third, a regression artificial neural network is introduced, mapping the input to the obtained reduced output parameter space.

\subsection{Data set definition, generation, and split}
\label{ssec:data}

The data set used to train and evaluate the ANNs consists of a sequence of input tensors and corresponding output tensors. The former resemble the energy distributions of the particle flux towards the target $f_\mathrm{Ar}(E)$ (i.e., Ar IED) as well as the Ti$_{1-x}$Al$_x$ target surface state, that is the stoichiometry $x$. The latter comprise the EADs of all species, that is Ti, Al, and Ar.

IEDs $f_\mathrm{Ar}[k]$ are analytically specified and discretized at the $k$-th energy bin $E_k$ similar to \cite{kruger_machine_2019}. In equal parts, the IED is chosen to be either a mono-energetic, a discrete Gaussian (standard deviation $\sigma=20$\,eV) or bimodal pulsed (width $\Delta_E = 40$\,eV). For all of which an energy grid size of 10 eV is used. The mean energies of the bombarding Ar ions are increased from 0\,eV to 1,500\,eV for the mentioned sequence of distribution functions. IEDs at the boundary (0, 1,500\,eV) are cropped for the sake of simplicity. This corresponds to a data set size per stoichiometry of 450. Stoichiometry $x \in \left\{ 0.3, 0.5, 0.7 \right\}$ is varied for training and evaluation, whereas $x=0$ signifies pure Ti. In addition, data sets with stoichiometry $x \in \left\{ 0.2, 0.4, 0.8 \right\}$ are selected exclusively for post-training investigation of the generalization capabilities of the obtained regression model (i.e., interpolation, extrapolation). These data sets were excluded from training. Thus, the training input parameter space is explored by a total of 1,350 samples for this case study. The data set is chosen to be small to mimic scenarios that are limited by computational resources (e.g., MD). The sputtering distributions are simulated with TRIDYN \cite{moller_tridyn_1984}, providing for each input the corresponding output tensor of three EADs $Y[i,j]$ with $E_i \in \left[0, 30 \right]$\,eV in 30 steps and $\cos(\vartheta_j) \in \left[ 0, 1 \right]$ in 20 steps, equidistantly. For the proceeding training, the data is padded with zeros to obtain data sizes corresponding to powers of 2 -- specifically an output data point $Y[i,j]$ has shape ($32 \times 24 \times 3$). In TRIDYN, any target material is assumed to be amorphous, with a given initial stoichiometry $x$ and a maximum Ar incorporation of 10 \%. Moreover, the collision cascade is described by the binary collision approximation. The latter is theoretically only valid for high kinetic energies ($\gtrsim$ 1\,keV), but TRIM based simulations have been shown to be also applicable in cases of lower energies \cite{behrisch_sputtering_2007}. Detailed descriptions and reviews of the TRIM method and its parameters can be found elsewhere \cite{biersack_monte_1980,eckstein_sputtering_1984, moller_tridyn_1984,behrisch_sputtering_2007, hofsass_simulation_2014}. The parameters used in this study are listed in the appendix (Tables~\ref{table:global tridyn_parameters} and \ref{table:element specific tridyn_parameters}). Two choices for the number of projectiles are chosen (i.e., $10^4$ and $10^6$). They are meant to resemble challenging statistical EAD representations, as possibly obtained by computationally more demanding methods for example MD, and estimating ``ground truth'' EADs, respectively. The absolute frequencies of the latter are divided by 100 to normalize the EADs for evaluation of the network performance with comparable absolute frequencies. The data has not been normalized or standardized in any other way. 

The data set is shuffled and then split into three subsets (i.e., 80 \% training,  10 \% validation, and 10 \% test set). The first is used to compute the loss by forward propagation of the input variables through the network, which is subsequently backward propagated. The error gradient is accumulated over the size of the chosen mini batch, 32. All trainable weights are then updated by a stochastic gradient descent algorithm, that is adaptive moment estimation (Adam) \cite{kingma_adam_2015}. After each epoch, the validation loss is used to decide whether the learning rate has to be reduced and eventually the training is stopped. The test set is finally used to evaluate the network's generalization capability and, for instance, to choose the best set of hyperparameters (HPs) in case of HP studies. Cross validation (CV) is considered to reduce the potential bias by how the data set is randomly distributed among its subsets. The validation set is consecutively interchanged with an equivalent sized fraction of the training set to yield a new combination. In case of three subsets, nested CV may be the most appropriate choice. However, the nested approach leads to a recognizable computational burden due to all the combinatorial possibilities (e.g., one obtains $K \times L$ evaluations for a single set of HPs and outer $K$-fold and inner $L$-fold CV). As a compromise, the total data set is consecutively shifted through fixed split markers during the CV, which means that in the end all data has been used only once for validation and testing over the course of a 10-fold CV.

For the final training with optimal HPs and challenging statistics, and for comparison with the estimated ground truth (full data set), the test set is evenly distributed among the training and validation subsets. To obtain an integer number of CV cases, we proceed with the same 10-fold CV and distribute the original 10~\% test set for each CV run (i.e., 85~\% training, 15~\% validation set with 5~\% overlap of the validation set for subsequent CV runs). By training/evaluating a corresponding ensemble of $K=10$ ANNs with identical HPs, the model configuration's total score as well as its variation and sensitivity to the presented input data and random initialization may be assessed (detailed later).

\subsection{Dimensionality reduction}
\label{ssec:Dimensionality_reduction}

The IEDs of impinging ions and the EADs of sputtered particles are described by histograms with appropriately sized bins. In consequence, a relatively large number of bins corresponding to input nodes ($151 + 1 = 152$ for the IED and $x$) as well as output nodes ($30 \times 20 \times 3 = 1800$ for the EADs of all species) are required. Setting up the targeted regression model by means of a MLP, as previously proposed \cite{kruger_machine_2019}, inherently leads to a complex network (high number of degrees of freedom). This may in particular become an issue when further reducing the data size and (or) the statistical accuracy. 

In contrast, a dimensionality reduction may be achieved through the concept of autoencoder neural networks \cite{hinton_reducing_2006}. Specifically, a $\beta$-variational autoencoder ($\beta$-VAE) is chosen to generalize trends in the data and reduce the dimensionality at the same time. VAE networks consist of an encoder, a reparameterization sampler, and a corresponding decoder \cite{kingma_auto-encoding_2013, rezende_stochastic_2014}. $\beta$-VAEs introduce the $\beta$ factor to disentangle the latent space representation \cite{higgins_beta-vae_2016, burgess_understanding_2017}. The encoder maps the EADs of all species to a low dimensional latent space representation with a corresponding number of mean and variance values. The latter are input to the sampler model and used to reparameterize a standard normal distribution. The obtained (pseudo-)random samples are fed into the decoder, which pursues to reconstruct the output signal. Passing the information through the latent space (bottle neck) favors the generalization process, which may be further enhanced by choosing simple encoder and decoder networks. 2D convolutional layers (CLs) are utilized as hidden layers in this work. Because of their shared parameters, CLs inherently require less trainable weights than fully-connected dense layers (DLs). This may also be understood by the correlation of neighboring EAD bins. In case of CLs, multiple convolutional kernels (also called channels or filters) are passed over the input, producing the output by applying the dot product. Herein, we stick to the convention that the data structures' last dimension represents the number of channels of the CL. The input field size is preserve by padding the data with zeros at the boundaries. The overall operation is meant to extract features from the input signal and provide them to subsequent layers. To obtain a nonlinear relation, the information is typically passed through an nonlinear activation function. By using a stride $s>1$, the data is sampled down (e.g., for $s=2$ the kernel skips every second input node effectively reducing the input field size by two).

\begin{figure}[t]
\includegraphics[width=8cm]{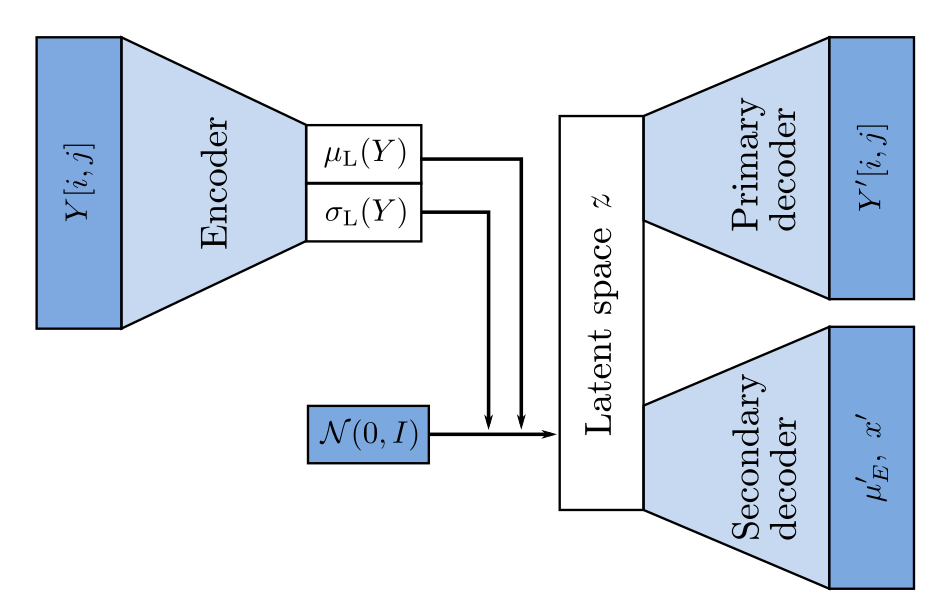}
\caption{Schematic of the VAE network structure.}
\label{fig:VAE}
\end{figure}

In the following, the utilized ANNs are described. If not stated differently, the stride is chosen to be $s=2$ for any hidden CL of the encoder as well as for upsampling deconvolution layers of the decoder, and $s=1$ for the output CLs, respectively. The activation functions for any hidden layer and output layers are set as rectified linear unit (ReLU) and linear, respectively, if not explicitly stated otherwise.

A schematic of the VAE network is depicted in Figure \ref{fig:VAE}, whereas a detailed graph is presented in the appendix. The encoder takes as input $Y[i,j]$, the stacked EADs with dimensions ($32 \times 24 \times 3$). It begins with a sequence of three 2D CLs with 6, 12 and 24 filters as well as ($5\times5$), ($3\times3$) and ($1\times1$) kernel sizes, respectively. The filter number is doubled to account for the skipped input nodes in both dimensions, which reduces the information volume by a factor of two (favoring generalization). The kernel sizes are chosen to initially extract coarse and later fine features, while also respecting the particular input width and height of each layer. The extracted features ($4\times3\times24$) are flattened (288) and interpreted by two dense output layers, each matching the latent space dimensions $n_\mathrm{l}$. The first of these DLs represents the array of means $\mu_\mathrm{L}(Y)$ and the second the array of standard deviations $\sigma_\mathrm{L}(Y)$ of $n_\mathrm{l}$ normal distributions.

Stochastic backpropagation is enabled via the so-called reparameterization trick \cite{kingma_auto-encoding_2013,higgins_beta-vae_2016}. Specifically, a dedicated (pseudo-)random sampling input layer is implemented, which samples from a standard normal distribution. Its samples are scaled by the standard deviations $\sigma_\mathrm{L}(Y)$ and shifted by the mean values $\mu_\mathrm{L}(Y)$, previously output by the encoder network. It therefore redefines the randomness as an input to the model.

The decoder is often chosen to be an equivalent counterpart to the encoder. However, here, an asymmetric $\beta$-VAE is found to be superior to a symmetric one. The structure and topology of the primary decoder is as follows: A single input layer takes $n_\mathrm{l}$ latent space coordinates (e.g., generated from the sampling layer). A subsequent DL with $288 c_\mathrm{a}$ nodes is used to establish the corresponding initial set of features. The hyperparameter $c_\mathrm{a}$ is meant to boost the asymmetry by enabling the decoder to set up more features than the encoder provides. Thereafter, the output is reshaped to match the width and height of the encoder's last CL output. Third, a sequence of three transposed 2D CLs with 24, 12 and 6 filters as well as ($3\times3$), ($5\times5$) and ($5\times5$) kernel sizes are used, respectively. As for the output CL, 3 filters (for 3 species) and a ($1\times1$) kernel is utilized in combination with the absolute value of the linear activation function. The last prohibits the occurrence of negative absolute frequencies in the predicted/reconstructed EADs $Y^\prime[i,j]$. Here and throughout the text, primes denote predictions by the neural network.

In addition, a secondary decoder is introduced. This decoder is meant to reconstruct the mean of the input IED $\mu_E^\prime$ and the stoichiometry $x^\prime$ from the latent space, and to condition the latter. The submodel consists of two hidden DLs with $n_\mathrm{id}$ nodes, and two output DLs, each with 1 node. 

Losses are defined by the Kullback-Leibler-Divergenz (KL) loss as a function of the encoder output \cite{kingma_auto-encoding_2013, rezende_stochastic_2014}, and the weighted mean squared error (MSE) as reconstruction loss function of both decoder outputs. The weights for the reconstruction of the EADs $Y^\prime$ and the stoichiometry $x^\prime$ are 1, whereas $10^{-6}$ is used for the mean of the IED $\mu_E^\prime$. The $\beta$ factor of the $\beta$-VAE is used to weight the KL loss with respect to the MSE, disentangling the latent space representation \cite{higgins_beta-vae_2016, burgess_understanding_2017}. Simulated annealing is used over the course of 100 epochs to gradually increase $\beta$ from 0 to its target value per batch. The latter is determined in a hyperparameter study (detailed in Section~\ref{sssec:results_dimensionality_reduction_hyperparameter_study}).

The coefficient of determination $R^2=1-\frac{SS_\mathrm{res}}{SS_\mathrm{tot}}$ is considered as an additional metric to judge the network's performance with $SS_\mathrm{res}$ and $SS_\mathrm{tot}$ being the sum of squared residuals and total sum of squares, respectively. A value of unity indicates fully explained residuals. The MSE and $R^2$ for multiple outputs are taken as their sum and average, respectively. Individual metrics are marked by the particular subscript (e.g., MSE$_\mathrm{EAD}$).

The learning rate is fixed at $10^{-3}$ during the simulated annealing of $\beta$. Afterwards, the learning rate is reduced by 2 whenever 
the validation loss does not improve on its previous minimum value by more than $10^{-3}$ (less than 0.1 \% of the typically observed loss) over the course of 10 epochs. When the validation loss does not improve by more than $10^{-3}$ for 25 epochs, early stopping terminates the training phase.

\subsection{Regression model}
\label{ssec:Regression_model}

\begin{figure}[t]
\includegraphics[width=8cm]{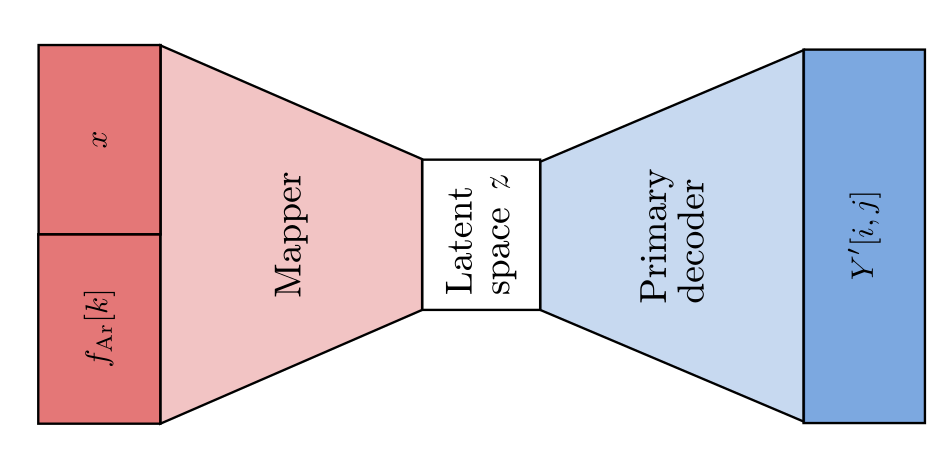}
\caption{Schematic of the regression network structure. The primary decoder is transferred from the previously trained $\beta$-VAE network and not trainable.}
\label{fig:regression_network}
\end{figure}

To finally establish a regression model, input variables need to be mapped to the VAE's latent space. Input for the regression are the discretized Ar IED $f_\mathrm{Ar}[k]$ and the stoichiometry $x$ of the surface composite. The decoder predicting the EADs $Y^\prime[i,j]$ is taken from the $\beta$-VAE, while its weights are set to be non-trainable. Thus, the learning progress from the preceding step is transferred. A schematic of the combined network is depicted in Figure \ref{fig:regression_network}.

The input $f_\mathrm{Ar}[k]$ is initially downsampled with a sequence of six 1D CLs with kernel sizes and strides of (3) and 2, respectively. The initial channel depth is chosen to be 1 (for a single species: Ar$^+$) and is consecutively doubled up to the $m$-th subsequent layer ($m$ being a hyperparameter), fully compensating the decreasing channel width by increasing its number. For the remaining layers, the number of filters is kept constant, halving the information volume per layer. The output features $o_\mathrm{f}$ are flattened, concatenated with the input $x$ and processed with a single hidden DL. The number of output nodes of this layer is selected as the scaled sum of the preceding and following layers' nodes, $c_\mathrm{s}(o_\mathrm{f}+1+n_\mathrm{l}$). The scaling factor $c_\mathrm{s}$ is taken into account as a hyperparameter. The final output of the mapper is implemented by a DL with $n_\mathrm{l}$ nodes and a linear activation function.

MSE and $R^2$ are used as loss function and additional metric, respectively. The learning rate and termination of the training process is maintained as described in the preceding Section~\ref{ssec:Dimensionality_reduction}. 

All ANNs are set up with the TensorFlow framework 2.4.1 and the Keras API included therein \cite{abadi_tensorflow:_2016, chollet_keras:_2015}.

%%%%%%%%%%%%%%%%%%%%%%%%%%%%%%%%%%%%%%%%%%%%%%%%%%%

\section{Results}
\label{sec:results}

This section begins by studying the optimization of the dimensionality reduction and is followed by the presentation as well as discussion of the latent space. Subsequently, different network configurations are studied to obtain the best regression model, whose predictions are eventually compared with the estimated ground truth.

\begin{table}
\begin{tabular}{lc}
\hline
hyperparameter & range  \\  
\hline
$n_\mathrm{l}$ & [1, 2, 3] \\
$\beta$ & [0.0, 0.1, 0.5, 1.0] \\
$c_\mathrm{a}$ & [1.0, 1.5, 2.0, 2.5, 3.0] \\
$n_\mathrm{id}$ & [2, 4, 8, 16] \\
\hline
$m$ & [0, 1, 2, 3, 4, 5, 6] \\
$c_\mathrm{s}$ & [0.5, 1.0, 1.5, 2.0] \\
L2 & [$0.0,~10^{-5},~10^{-4},~10^{-3},~10^{-2}$] \\
\hline
\end{tabular}
\caption{Parameter range for HP studies of the $\beta$-VAE and the regression ANN to optimize dimensionality reduction and generalization. HPs are described in Sections~\ref{ssec:Dimensionality_reduction} and \ref{ssec:Regression_model}.}
\label{tab:HPcVAE}
\end{table}

\subsection{Dimensionality reduction and latent space representation}
\label{ssec:results_dimensionality_reduction}

\subsubsection{Hyperparameter study}
\label{sssec:results_dimensionality_reduction_hyperparameter_study}

A convolutional $\beta$-VAE is employed in a HP study to set up the (presumably) most optimal network for the reduction of the EAD parameter space to the low-dimensional latent space. To obtain the best network structure for the given task, a variation of the latent space dimensions $n_\mathrm{l}$, KL loss factor $\beta$, asymmetry boost factor $c_\mathrm{a}$ for the primary decoder and number of nodes $n_\mathrm{id}$ per hidden layer of the secondary decoder is investigated.

The total range of HPs are listed in Table \ref{tab:HPcVAE}. For computational exploration, the complete HP space is iterated over in a nested grid-based scheme. To facilitate an intuitive interpretation of the results, however, the discussion is restricted to only the best combinations given a respective HP to be varied. These are examined one after another. The remaining HPs are chosen to minimize a combined metric (CM),
\begin{align}
\label{eq:cm}    
\mathrm{CM}=\mu_\mathrm{MSE}+\sigma_\mathrm{MSE}+c_{R^2}(1-\mu_{R^2}+\sigma_{R^2}),
\end{align}
with the MSE mean $\mu_\mathrm{MSE}$, MSE standard deviation $\sigma_\mathrm{MSE}$, $R^2$ mean $\mu_{R^2}$, $R^2$ standard deviation $\sigma_{R^2}$ for the 10-fold CV, and the factor $c_{R^2}$ being chosen as 4 to account for the different scales of the metrics. This provides an upper bound for the CM. The contribution of the KL loss MSE$_\mathrm{KL}$ is neglected with respect to the reconstruction loss, when evaluating the model performances. 

For all figures in this section, $\mu_\mathrm{MSE}\pm\sigma_\mathrm{MSE}$ and $\mu_{R^2}\pm\sigma_{R^2}$ are displayed by the particular error bars. Transparent regions are meant to guide the eye.

\begin{figure}[t]
\subfloat[Latent space dimension $n_\mathrm{l}$.\label{fig:HPStudy_lSpace}]{%
    \includegraphics[width=8cm]{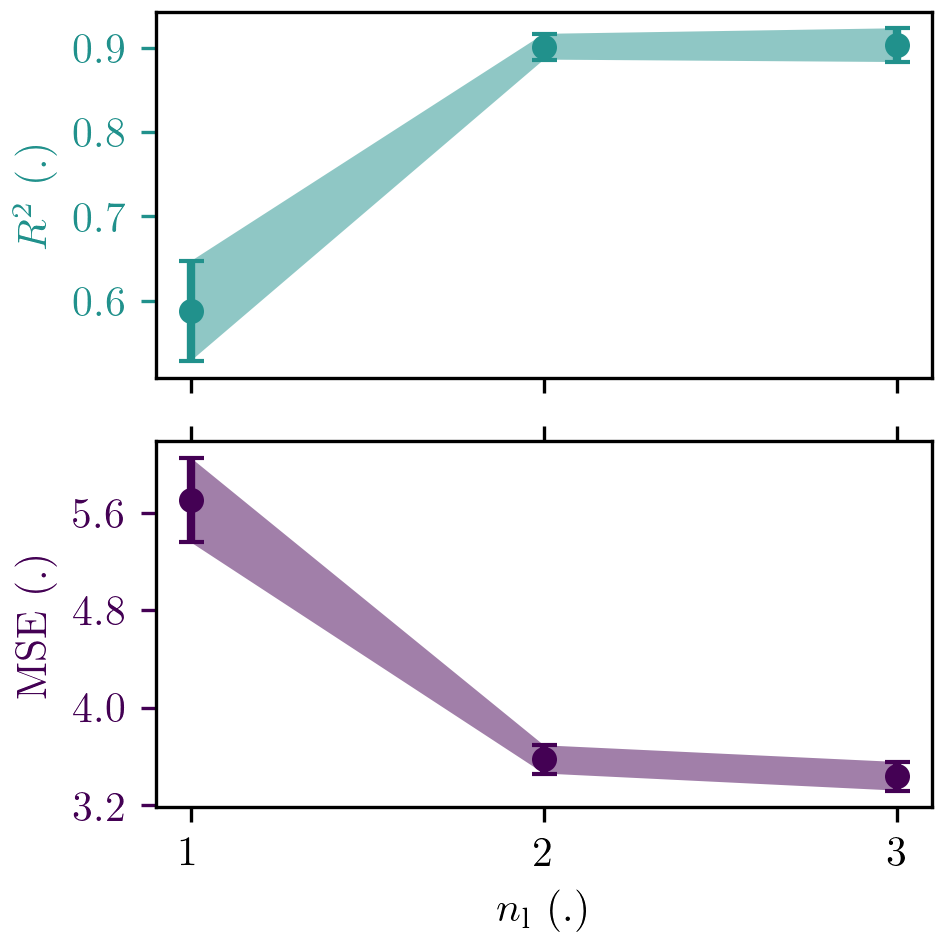}
}\hfill
\subfloat[KL loss factor $\beta$.\label{fig:HPStudy_klCoeff}]{%
    \includegraphics[width=8cm]{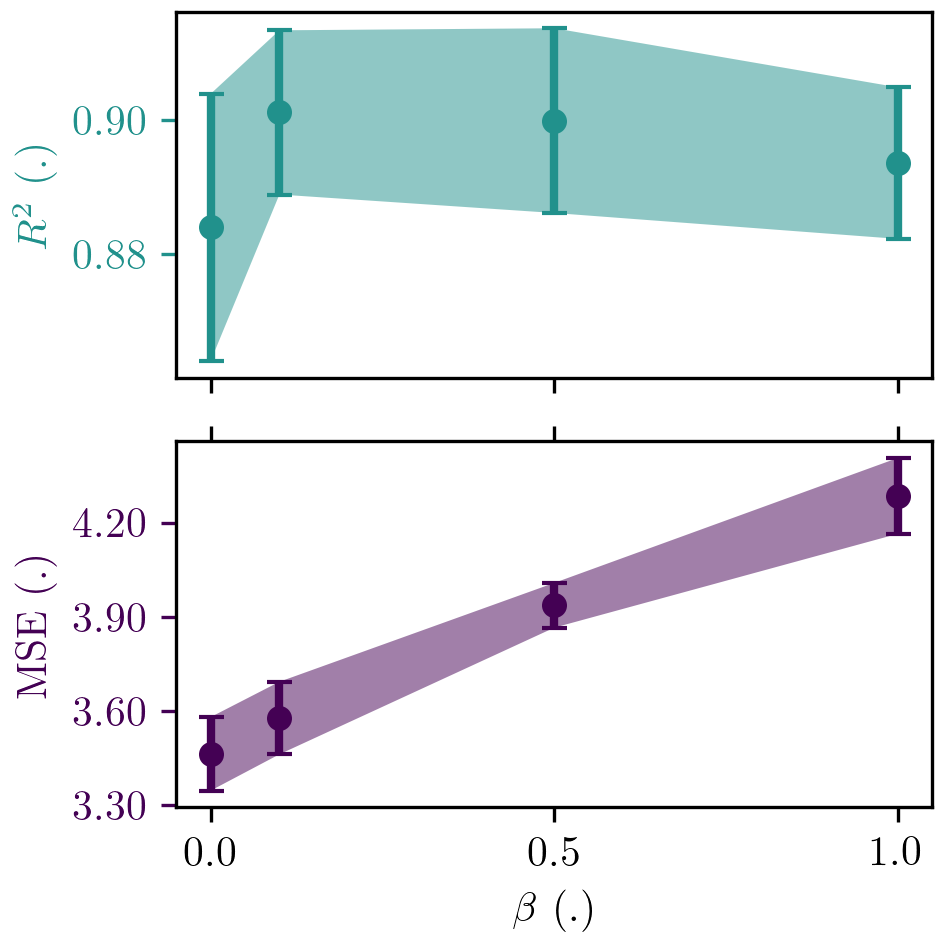}
}\hfill
\subfloat[Asymmetry boost factor $c_\mathrm{a}$.\label{fig:HPStudy_asyms}]{%
    \includegraphics[width=8cm]{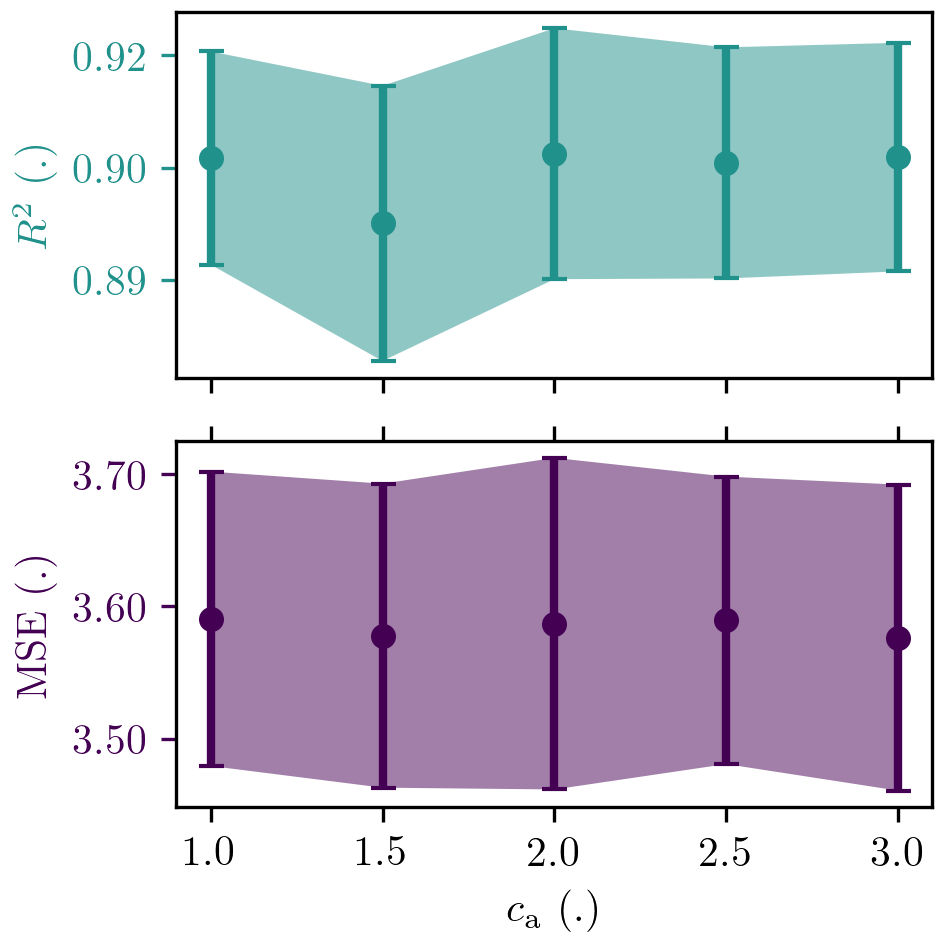}
}\hfill
\subfloat[Number of nodes $n_\mathrm{id}$ for the secondary decoder.\label{fig:HPStudy_idNodes}]{%
    \includegraphics[width=8cm]{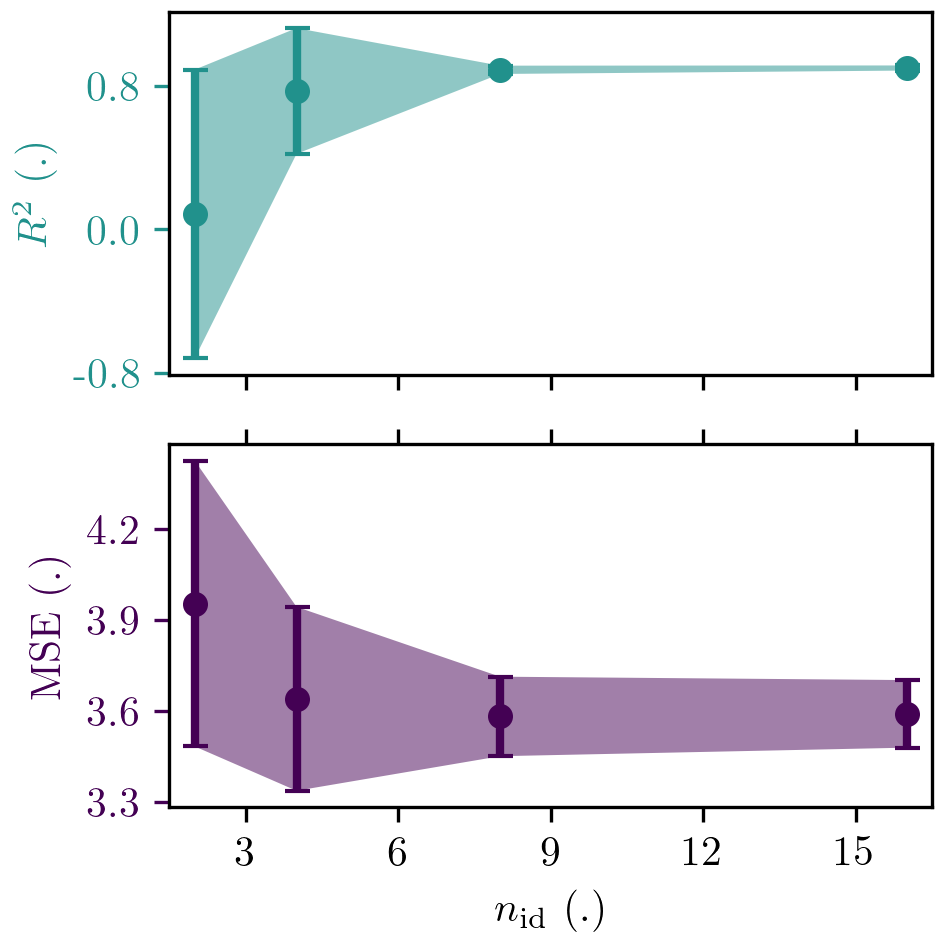}
}
\caption{HP optimization for the $\beta$-VAE. Other HPs are chosen to minimize the CM defined in Eq. \eqref{eq:cm}. The MSE and $R^2$ error bars equal $\mu_\mathrm{MSE}\pm\sigma_\mathrm{MSE}$ and $\mu_{R^2}\pm\sigma_{R^2}$, respectively.}
\end{figure}

In Figure \ref{fig:HPStudy_lSpace}, the MSE as well as the $R^2$ metric are shown for HP sets \{1, 1.0, 2.0, 16\}, \{2, 0.1, 3.0, 16\} and \{3, 0.0, 1.0, 16\} as \{$n_\mathrm{l}$, $\beta$, $c_\mathrm{a}$, $n_\mathrm{id}$\}. As apparent from the variation of $n_\mathrm{l} \in \{1,2,3\}$, a 1D latent space is too restricted to allow for a reasonable reconstruction. Although $n_\mathrm{l}=3$ improves the overall VAE performance slightly, the 2D latent space is favored to obtain a more reduced parameter space and, hence, a simpler regression problem. This is signified by the mean value of $R^2$ ($\mu_{R^2}=0.901$) as well as its standard deviation ($\sigma_{R^2}=0.015$) and indicates that the model has learned to generalize successfully.

Second, the KL loss factor $\beta \in \{0.0, 0.1, 0.5, 1.0\}$ is varied, while $n_\mathrm{l}=2$ is kept constant. The results are displayed in Figure~\ref{fig:HPStudy_klCoeff} for HP sets \{0.0, 3.0, 16\}, \{0.1, 3.0, 16\}, \{0.5, 1.0, 16\} and \{1.0, 3.0, 16\} as \{$\beta$, $c_\mathrm{a}$, $n_\mathrm{id}$\}. The $\beta$-VAE reduces to an ordinary AE for $\beta=0$. This case provides the smallest reconstruction loss (MSE) at the cost of the lowest $R^2$ mean and highest $R^2$ standard deviation. Increasing $\beta$ to 0.1 slightly worsens the reconstruction loss, but yields an important improvement regarding the coefficient of determination. This emphasizes the impact of the statistical variation on the training success. Further increments of $\beta$ to 0.5 or 1.0 lead to worse-performing networks. Hence, $\beta=0.1$ is considered here to resemble the best trade off between $R^2$ and MSE.

In the following, the latent space dimension $n_\mathrm{l}=2$ and the KL loss factor $\beta=0.1$ are maintained. In Figure~\ref{fig:HPStudy_asyms} the variation of the asymmetry boost factor $c_\mathrm{a} \in \{1.0, 1.5, 2.0, 2.5, 3.0\}$ for the primary decoder is shown for $n_\mathrm{id} = \{16, 8, 16, 16, 16 \}$. The performance is relatively insensitive with regard to this quantity. Hence, $c_\mathrm{a} = 1.0$ is chosen to minimize network complexity.

At last, for $n_\mathrm{l}=2$, $\beta=0.1$, and $c_\mathrm{a}=1.5$, the number of nodes $n_\mathrm{id}$ per hidden layer of the secondary decoder are varied. The corresponding MSE and $R^2$ metrics are presented in Figure~\ref{fig:HPStudy_idNodes}. A minimum network complexity is apparently required, that is $n_\mathrm{id} \geq 8$. Further increasing its value to 16 provides a competing and even better performance (i.e., $\mu_\mathrm{MSE}=0.901$, $\sigma_\mathrm{MSE}=0.014$) in comparison to 8 (i.e., $\mu_\mathrm{MSE}=0.891$, $\sigma_\mathrm{MSE}=0.023$).

\begin{table}
\begin{tabular}{lcccc}
\hline
metric & $\mu$ (.) & $\sigma$ (.) & $\mu_\mathrm{g}$ (.) & $\sigma_\mathrm{g}$ (.) \\  
\hline
$\mathrm{MSE}$ 		 		& 3.591 & 0.111 & 0.226 & 0.119 \\
$\mathrm{MSE_{EAD}}$ 		& 3.580 & 0.109 & 0.217 & 0.116 \\
$\mathrm{MSE}_{\mu_E}$ 	 	& 0.009 & 0.002 & 0.008 & 0.003 \\
$\mathrm{MSE}_x$ 			& 0.001 & 0.000 & 0.001 & 0.000 \\
$\mathrm{MSE}_\mathrm{KL}$  & 0.455 & 0.008 & 0.452 & 0.014 \\
$R^2$ 						& 0.901 & 0.014 & 0.970 & 0.011 \\
$R^2_\mathrm{EAD}$ 			& 0.798 & 0.017 & 0.985 & 0.008 \\
$R^2_{\mu_E}$ 				& 0.949 & 0.006 & 0.955 & 0.017 \\
$R^2_x$ 					& 0.957 & 0.020 & 0.968 & 0.008 \\
\hline
\end{tabular}
\caption{MSE and $R^2$ metrics for the $\beta$-VAE, HPs being $n_\mathrm{l}=2$, $\beta=0.1$, $c_\mathrm{a}=1.0$ and $n_\mathrm{id}=16$. Metrics with subscript ``g'' are evaluated against the estimated ground truth data for which the test sets have been evenly distributed among the training and validation set as detailed in Section~\ref{ssec:data}.}
\label{tab:HPcVAE_performance}
\end{table}

The final set of HPs ($n_\mathrm{l}=2$, $\beta=0.1$, $c_\mathrm{a}=1.0$, $n_\mathrm{id}=16$) corresponds to 2,584 trainable weights for the encoder, 15,111 for the primary decoder and 354 for the secondary decoder. Thus, the $\beta$-VAE is built with 18,049 degrees of freedom. The individual contributions to the MSE and $R^2$ are listed in Table~\ref{tab:HPcVAE_performance} and indicate a reasonable reconstruction of the EADs. However, a reliable assessment is limited by the statistics of the test data. A well-generalizing prediction (which is expected to be comparable to the ground truth) intrinsically deviates from the test data in terms of the excluded erroneous fluctuations. Consequently, the corresponding noise level defines the limits of the obtained metrics, specifically $\mathrm{MSE_{EAD}} \approx 3.6$ and $R^2_\mathrm{EAD} \approx 0.8$. This statistical uncertainty is solely contained in the EADs, not the IED or the stoichiometry.

In particular, the secondary decoder is determined to reliably interpret the latent space as indicated by the respective individual metrics (e.g., $R^2_{\mu_E} \approx R^2_x \approx 0.95$). This property will be used in the proceeding analysis to interpret the obtained low-dimensional latent space. A detailed analysis of the attainable metrics based on the estimated ground truth is provided in Section~\ref{sssec:results_regression_model_ground_truth}, whereas the relevant values are already contained in Table~\ref{tab:HPcVAE_performance}.

\begin{figure}[t]
\includegraphics[width=8cm]{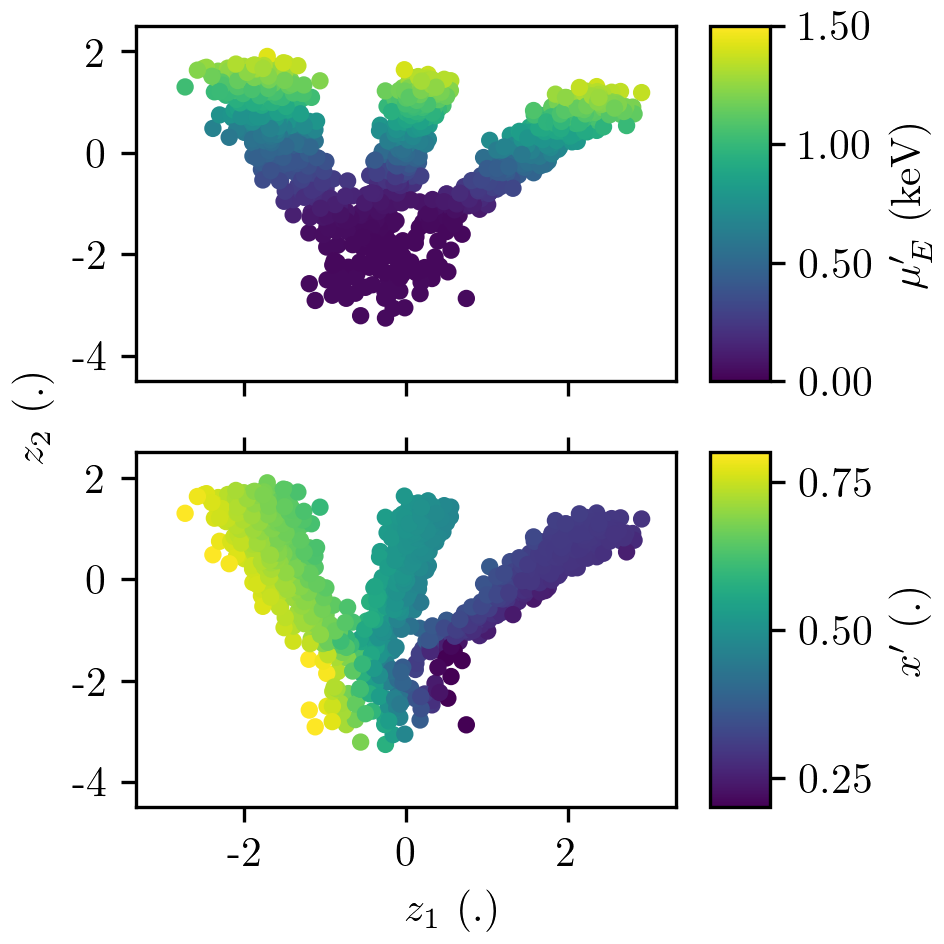}
\caption{Latent space ($z_1, z_2$) representation of the encoded EADs of sputtered species using the $\beta$-VAE. The stoichiometry $x^\prime$ and the mean of the IED $\mu_E^\prime$ are reconstructed by the secondary decoder.}
\label{fig:latent_space_VAE}
\end{figure}

\subsubsection{Latent space}
\label{sssec:results_dimensionality_reduction_latent_space}

In this section, the low-dimensional latent space representation is studied for the complete data set (with challenging statistics). This analysis utilizes the optimal HPs as determined in the previous section. Depicted in Figure \ref{fig:latent_space_VAE} is the 2D latent space with coordinates $(z_1, z_2)$ for $n_\mathrm{l}=2$ dimensions. The plotted samples correspond to the individual data points, whereas the distribution of samples corresponds to the encoder projection after training, and the color coding is the corresponding interpretation of the secondary decoder ($\mu_E^\prime, x^\prime$). A number of aspects are noteworthy regarding its interpretation:
(1) The displayed latent space corresponds to what is input to both decoder networks, that is, compressed by the encoder and statistically evaluated by the sampling layer using the $n_\mathrm{l}$ mean values $\mu_\mathrm{L}(Y)$ and standard deviations $\sigma_\mathrm{L}(Y)$. Therefore, the given markers entail the respective statistical variation.
(2) This aspect is intrinsic to VAE neural networks and is the reason for their generalization properties. The latent space is effectively explored in between the original samples with the obtained standard deviations $\sigma_\mathrm{L}(Y)$. Hence, this facilitates improved interpolation capabilities in between training samples. It is for instance seen in the lower plot of Figure \ref{fig:latent_space_VAE} presenting the reconstructed stoichiometry $x^\prime$. While training is restricted to $x\in\{0.3, 0.5, 0.7\}$, the secondary decoder correctly identifies intermediate values.
(3) The KL divergence loss imposes a constraint to attract samples toward a standard normal distribution. Consequently, the $\beta$ scaling factor balances between making the latent space more compact and giving it the freedom to adapt to the reconstruction loss. $\beta=0$ corresponds to a pure reconstruction loss effectively eliminating any latent space regularization.
(4) Irrespective of the choice of $\beta$, it is noticeable that the obtained representation mainly distinguishes a varying stoichiometry at large mean ion energies $\mu_\mathrm{E}^\prime$ (as seen from the upper plot of Figure \ref{fig:latent_space_VAE}). At low energies the different cases overlap and are no more separable. This is even the case when considering only the latent distributions' mean values $\mu_\mathrm{L}(Y)$ (i.e., ignoring $\sigma_\mathrm{L}$) for all data points (not shown), which present narrow lines that ambiguously overlap at low energies. It suggests excellent generalization properties at the cost of a limited degree of certainty and is typical for the proposed $\beta$-VAE. This is an aspect that is revisited in Section \ref{sssec:results_regression_model_latent_space}.

\subsection{Regression model}
\label{ssec:results_regression_model}

The final regression problem to obtain a mapping between the inputs $f_\mathrm{Ar}[k]$ and $x$ to the output $Y^\prime[i,j]$ is drastically simplified by the dimensionality reduction, provided by the $\beta$-VAE. Specifically, $f_\mathrm{Ar}[k]$ and $x$ only have to be mapped to the latent space $z$, discussed in detail in the preceding section. The corresponding mapping from latent space $z$ to output EADs $Y^\prime[i,j]$ is predefined with the previously trained primary decoder, which is transferred while its weights are set non-trainable. The training of the mapper network is again optimized through a HP study as discussed initially. In the following, the mapping to the latent space is then discussed with respect to the trained representation of the $\beta$-VAE model. Finally, the regression model is completed and its prediction is compared to the estimated ground truth.

\subsubsection{Hyperparameter study}
\label{sssec:results_regression_model_hyperparameter_study}

To obtain the most appropriate ANN for the present problem, a variation of the network complexity (i.e., $m$ and $c_\mathrm{s}$) is considered. The number of channels of consecutive CLs are doubled up to the $m$-th layer and kept constant afterwards. The factor $c_\mathrm{s}$ is meant to adjust the number of nodes for the DL as a function of extracted features and latent space dimensions. $m \in \{0, 1, 2, 3, 4, 5, 6\}$ and $c_\mathrm{s} \in \{0.5, 1.0, 2.0\}$ are varied, respectively. Both quantities are described in detail in Section~\ref{ssec:Regression_model}. The same procedure as outlined in Section~\ref{sssec:results_dimensionality_reduction_hyperparameter_study} is applied and a single HP variation after another is examined, while the remaining HPs are selected to minimize the CM defined in Eq.~\eqref{eq:cm} over the course of a 10-fold CV. Also for the following figures, $\mu_\mathrm{MSE}\pm\sigma_\mathrm{MSE}$ and $\mu_{R^2}\pm\sigma_{R^2}$ are presented by error bars. Transparent regions are meant to guide the eye.

\begin{figure}[t]
\subfloat[$m$-th convolutional layer.\label{fig:HPStudy_Mfulllayers}]{%
    \includegraphics[width=8cm]{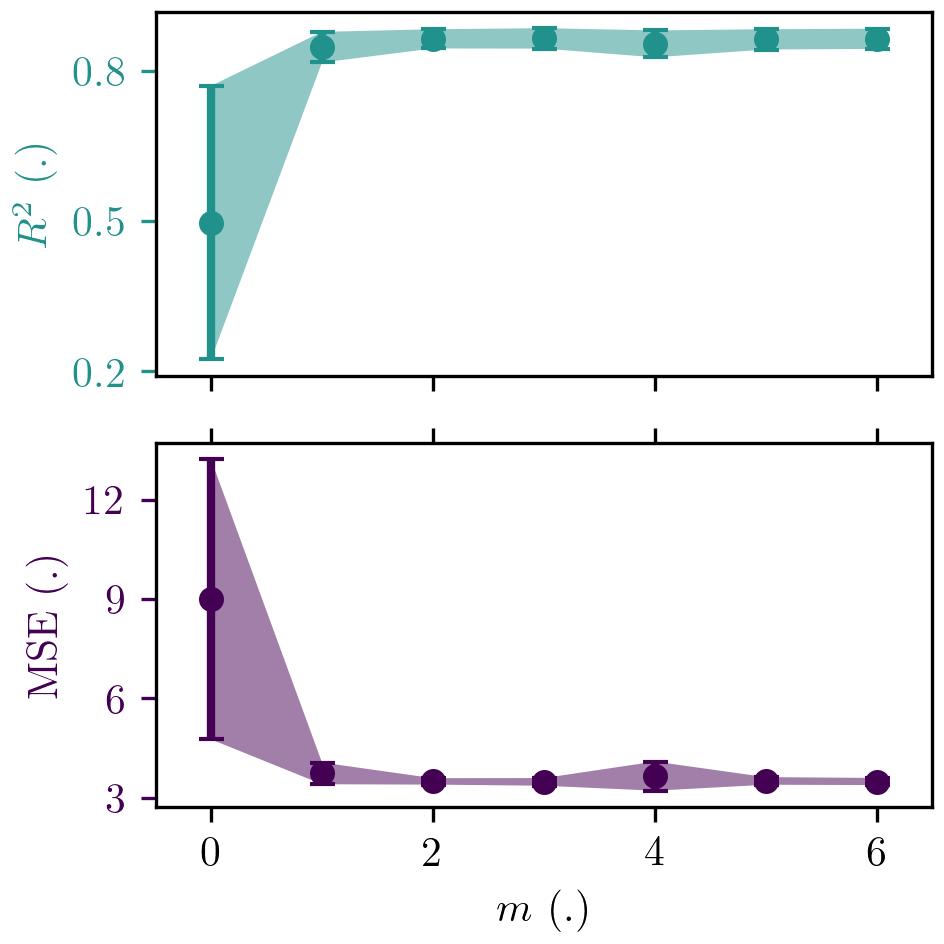}
}\hfill
\subfloat[$c_\mathrm{s}$ factor (hidden dense layer).\label{fig:HPStudy_MInterpreterScale}]{%
    \includegraphics[width=8cm]{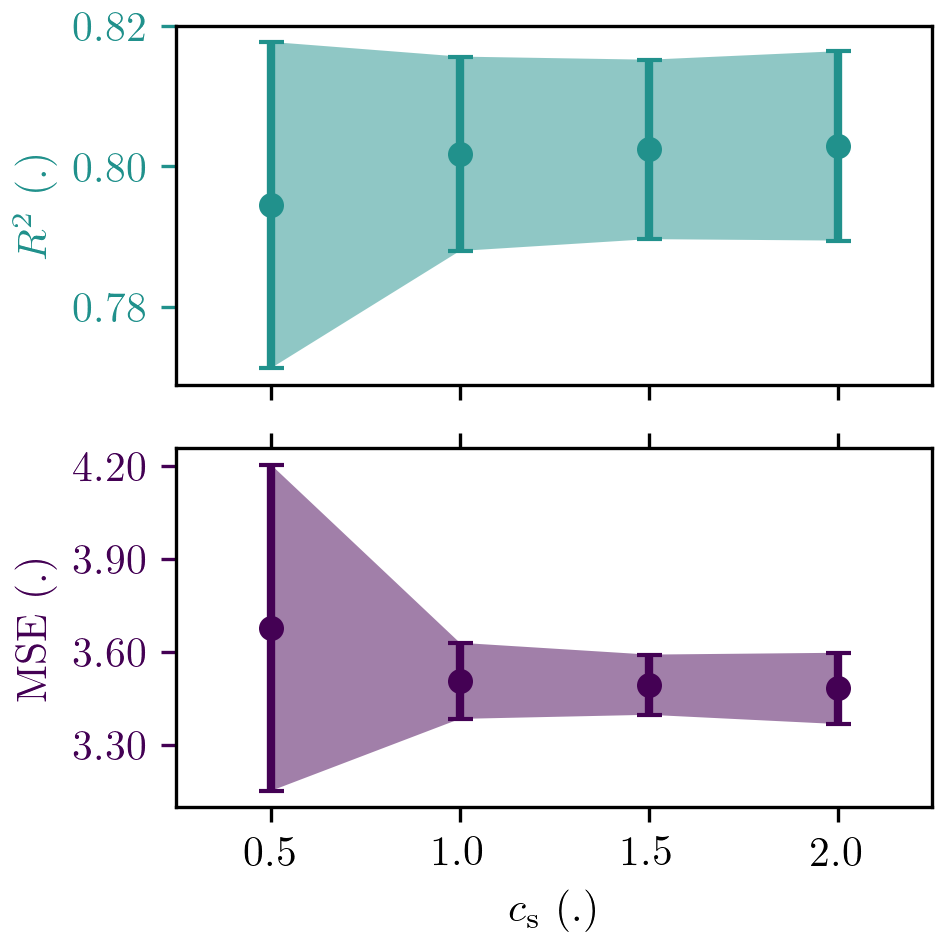}
}\hfill
\caption{HP optimization for the regression model. Other HPs are chosen to minimize the CM defined in Eq. \eqref{eq:cm}. The MSE and $R^2$ error bars equal $\mu_\mathrm{MSE}\pm\sigma_\mathrm{MSE}$ and $\mu_{R^2}\pm\sigma_{R^2}$, respectively.}
\end{figure}

In Figure~\ref{fig:HPStudy_Mfulllayers}, the MSE and $R^2$ metric are presented for \{0, 2.0\}, \{1, 1.0\}, \{2, 1.5\}, \{3, 2.0\}, \{4, 1.5\}, \{5, 0.5\} and \{6, 1.0\} as \{$m$, $c_\mathrm{s}$\}. The minimal complexity for a reasonable nonlinear projection is satisfied for $m=1$, whereas $m=2$ is considered as the optimal value. This means that the number of channels starting at a value of 1 is consecutively doubled up to the second CL and remains constant with a value of 4 for all remaining CLs. This can be reasoned by the balance between feature extraction and data compression through convolution: With kernel size (3) and stride 2, after the second CL already an energy range of 70\,eV of the initial input field is covered -- locally integrating the information of 7 bins, due to the respective overlap. It corresponds closely to the predefined features of the IED $f_\mathrm{Ar}[k]$, which span about 80\,eV  maximum in comparison. Therefore, subsequent CLs merely compress the extracted data requiring no additional compensation of the information volume.

Second, the $c_\mathrm{s}$ factor is varied for $m=2$. The resultant metrics are displayed in Figure~\ref{fig:HPStudy_MInterpreterScale} and reveal that the simplest hidden DL considered (i.e., $c_\mathrm{s}$=0.5) is insufficient for the targeted regression task. However, $c_\mathrm{s} \in \{1.0,1.5,2.0\}$ yield a similar performance, so that the least complex network (i.e., $c_\mathrm{s}=1$) is chosen. This means that the number of nodes for the hidden DL is the average of the nodes of the preceding (12 extracted IED features and 1 stoichiometry) as well as following (2 latent space dimensions) layer, resulting in 15 nodes. The mapper network is consequently constructed with 486 trainable weights for the final set of HPs. The corresponding mean and standard deviation of the $R^2_\mathrm{EAD}$ metric are $\mu_{R^2} = 0.802$ and $\sigma_{R^2} = 0.017$ as well as $\mu_\mathrm{MSE} = 3.506$ and $\sigma_\mathrm{MSE} = 0.123$ for the MSE$_\mathrm{EAD}$. While this again hints to a successful regression, a definite conclusion cannot be drawn as the obtained metrics are limited by the statistical quality of the test data (cf.\ Section~\ref{sssec:results_dimensionality_reduction_hyperparameter_study}). This aspect will be displayed more clearly in the context of the ground truth metrics presented in Section~\ref{sssec:results_regression_model_ground_truth}.

\subsubsection{Latent space}
\label{sssec:results_regression_model_latent_space}

\begin{figure}[t]
\begin{center}
\resizebox{8cm}{!}{
\includegraphics[width=8cm]{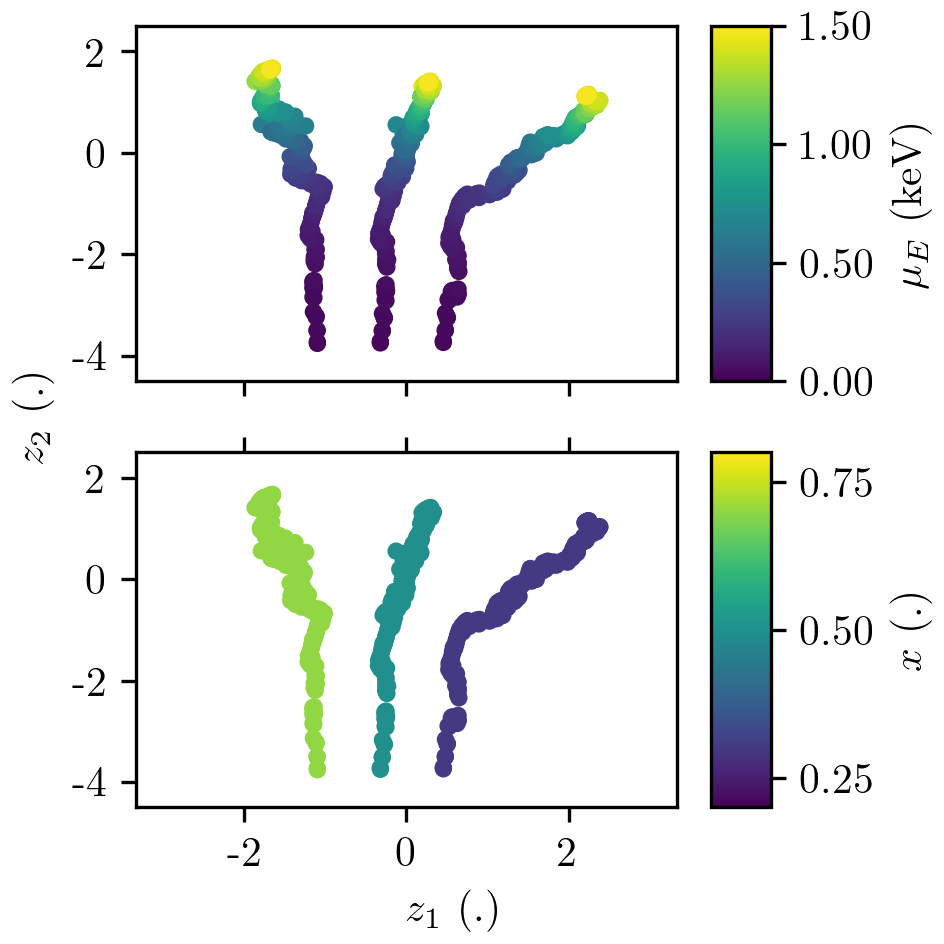}
}
\caption{Latent space ($z_1, z_2$) representation of the mapped EADs.}
\label{fig:latent_space_mapper}
\end{center}
\end{figure}

An additional indication of the training success can be obtained from an analysis of the latent space that the mapper network projects into. Again the complete data set (with the challenging statistical quality used for training) is assessed, whereas the input samples ($f_\mathrm{Ar}[k]$ and $x$) are directly projected to $(z_1, z_2)$ coordinates. Figure \ref{fig:latent_space_mapper} shows the corresponding latent space distribution with the IED mean $\mu_E$ and stoichiometry $x$ as color indication. As neither the input data nor the projection procedure is subject to any statistical variation, the definite input values and not the interpretation from the secondary decoder ($\mu_E^\prime$, $x^\prime$) are used for color coding. Notably, all different stoichiometries are individually mapped into narrow regions, well-discriminated from one another. From a comparison with Figure \ref{fig:latent_space_VAE}, their distribution over the latent space and the corresponding color coding (for both $\mu_E$ and $x$) closely follows the distribution obtained from the $\beta$-VAE. However, a striking difference is observed for low mean ion energies $\mu_E$. In this region the mapper is able to provide a clear separation for a varying stoichiometry, in stark contrast to the $\beta$-VAE. Due to this separation property and the circumstance that the mapper can adapt to the laid out latent space distribution, the complete regression network is so effective in the prediction of EADs for differing stoichiometry. This is further detailed in the subsequent section.

\subsection{Comparison with the estimated ground truth}
\label{sssec:results_regression_model_ground_truth}

To verify the hypotheses previously stated, the $\beta$-VAE and the regression ANN performance and their generalization properties can be evaluated by comparison with the estimated ground truth. As described at the end of Section~\ref{ssec:data}, the quality of prediction is quantitatively assessed through an ensemble of ANNs with identical HPs. At this point, however, their MSE and $R^2$ mean values and standard deviations are obtained with reference to the estimated ground truth data set. Specifically, the complete data set with a total of 1,350 data samples is used to score the final ANNs. Notably, the ground truth data is not required for the training procedure, but its metrics are merely presented here to quantitatively illustrate the success of our proposed approach.

For the $\beta$-VAE, a significant difference between the metrics with challenging statistical quality and the estimated ground truth metrics included in Table~\ref{tab:HPcVAE_performance} is immanent. The MSE$_\mathrm{EAD}$ mean value $\mu_\mathrm{MSE} = 0.217$ is more than an order of magnitude smaller, while the corresponding $\mu_{R^2_\mathrm{EAD}}$ value $\mu_{R^2} = 0.985$ approaches unity. This confirms the previous hypothesis that already the $\beta$-VAE generalizes well, with a performance comparable to the estimated ground truth. Consequently, the statistical noise inherent to the training data is successfully mitigated. This property is most relevant when only limited statistical quality data is attainable. It is exploited by the regression model which utilizes its pretrained decoder network.

The particular values for the regression model (i.e., combined mapper and decoder networks with fixed weights) are $\mu_\mathrm{MSE} = 0.125$ and $\sigma_\mathrm{MSE} = 0.059$ for MSE$_\mathrm{EAD}$ as well as $\mu_{R^2} = 0.991$ and $\sigma_{R^2} = 0.004$ for $R^2_\mathrm{EAD}$. Again the MSE$_\mathrm{EAD}$ is significantly lowered compared to its counterpart with challenging statistical quality, whereas $R^2_\mathrm{EAD}$ is improved and approaches unity. This signifies that the complete model has indeed learned to differentiate between noise and relevant physical features. The latter are effectively extracted and made available by the regression model. To provide an estimate of the computational effort required for inference, the evaluation time has been measured on a laptop computer equipped with an Intel i7-10510U CPU. The mean value and the standard deviation of the ANN prediction time for a single EAD is 0.310 ms and 0.002 ms, respectively.

\begin{figure}[t]
\begin{center}
\resizebox{16cm}{!}{
\includegraphics[width=16cm]{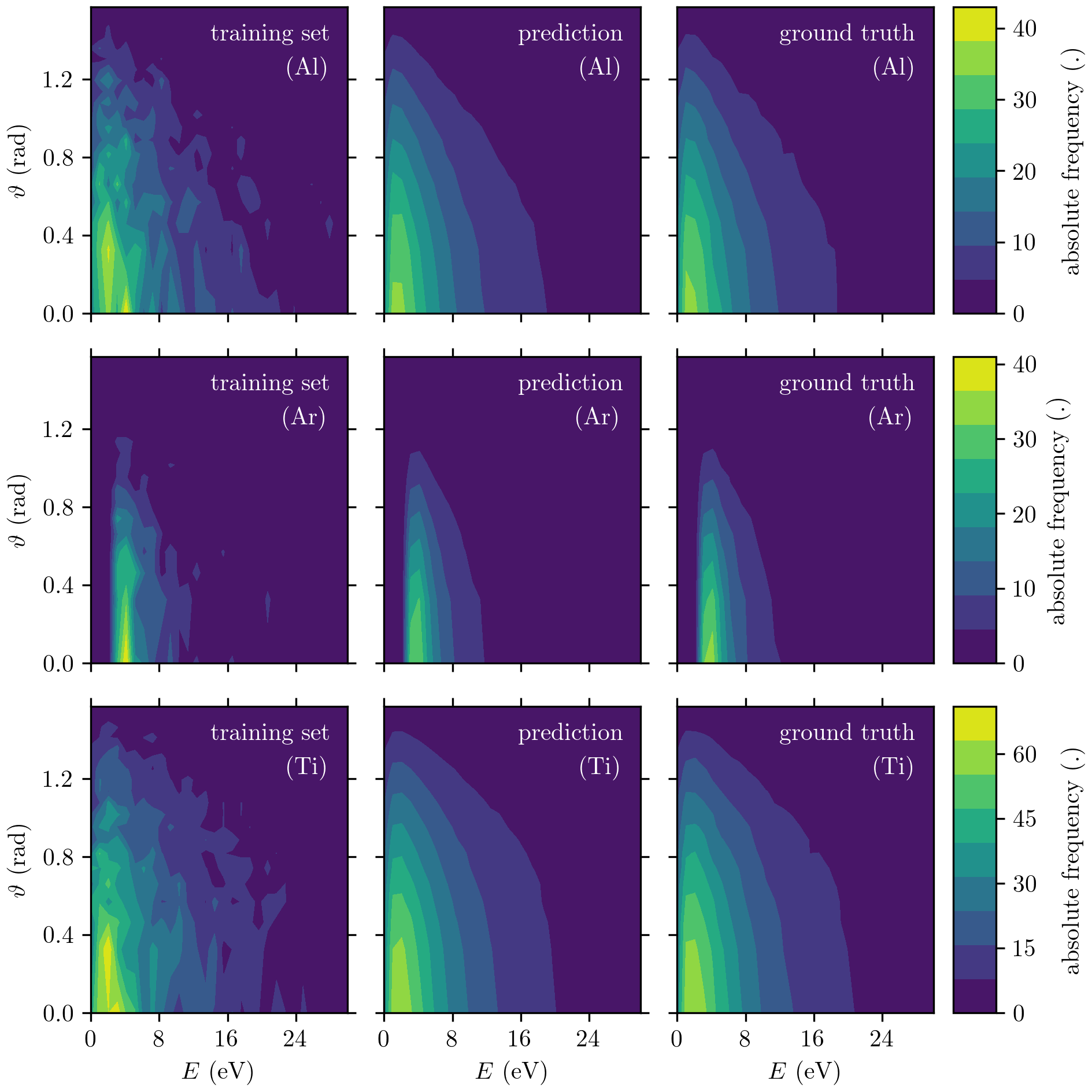}
}
\caption{Energy angle distributions per sputtered species (i.e., Al, Ar and Ti) for a Gaussian Ar IED with a mean of 1090\,eV and standard deviation of 20\,eV bombarding a Ti$_{0.7}$Al$_{0.3}$ composite target. The instance of the training set with its challenging statistics, the prediction of the regression model as well as the estimated ground truth are displayed.}
\label{fig:EAD_example_case}
\end{center}
\end{figure}

In the following, the prediction quality is further assessed in detail for an exemplary case, specifically a Gaussian Ar IED with a mean of 1090\,eV and standard deviation of 20\,eV bombarding a Ti$_{0.7}$Al$_{0.3}$ composite target. The corresponding EADs for all sputtered species (i.e., Al, Ar and Ti) with i) the challenging statistical representation, ii) the prediction of the ANN, and iii) the estimated ground truth are shown in Figure~\ref{fig:EAD_example_case}. The network clearly predicts the correct output distribution, which can be hardly distinguished from the ground truth, despite the noisy training reference. Smooth distributions are obtained, whereas characteristic features like the lack of Ar atoms with energies less than a few eV, or the energy peak of the Sigmund-Thompson energy distribution are contained and preserved. 

\begin{figure}[t]
\subfloat[Energy distribution.\label{fig:ED_example_case}]{%
    \includegraphics[width=8cm]{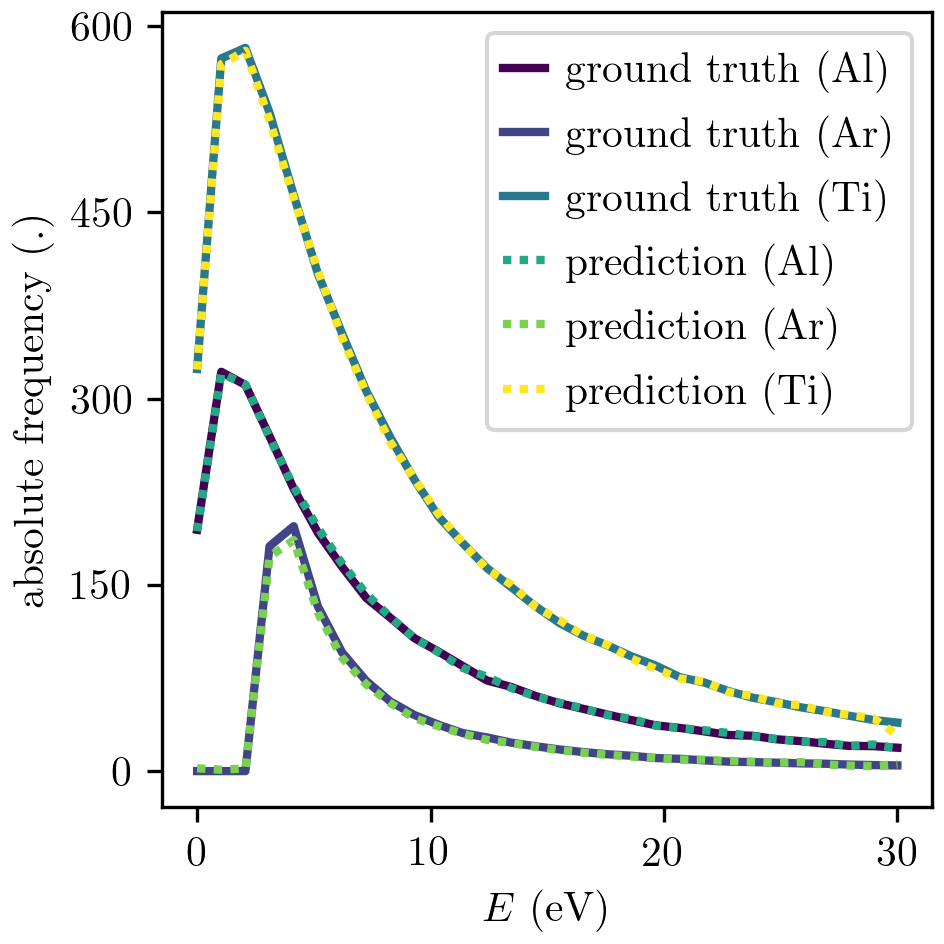}
}
\subfloat[Angle distributions.\label{fig:AD_example_case}]{%
    \includegraphics[width=8cm]{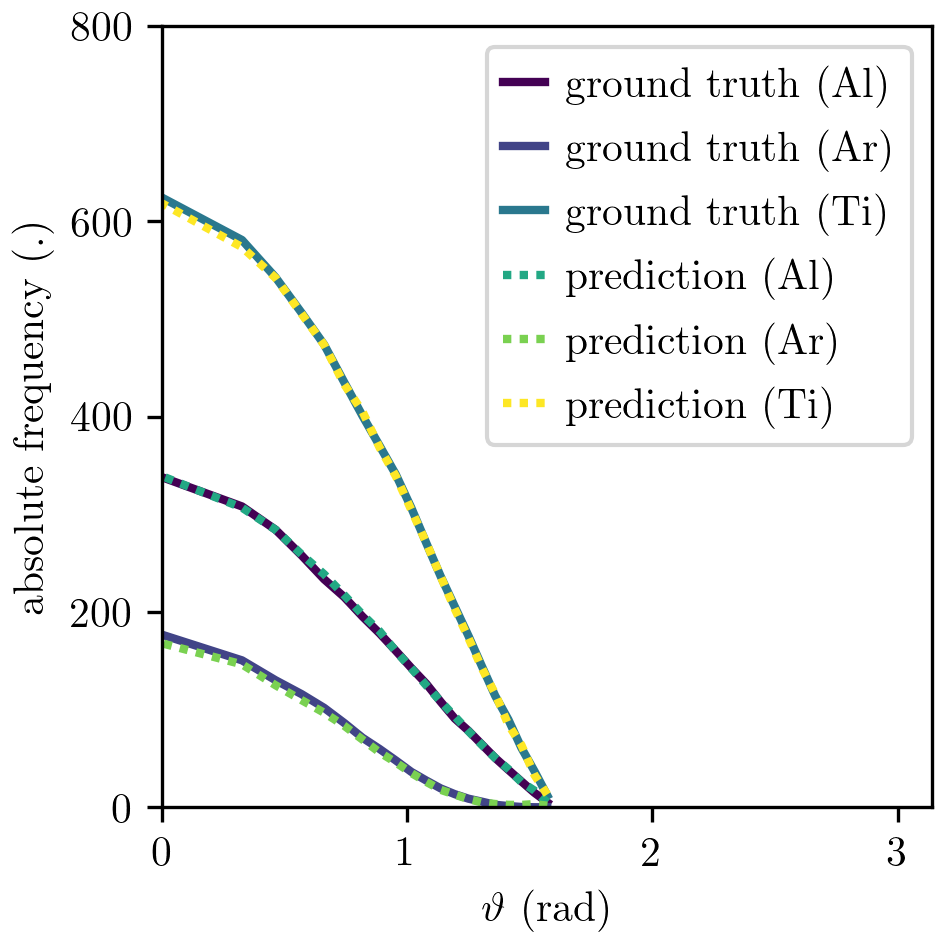}
}\hfill
\caption{Energy and angle distributions per sputtered species (i.e., Al, Ar and Ti) for a Gaussian Ar IED with a mean of 1090\,eV and standard deviation of 20\,eV bombarding a Ti$_{0.7}$Al$_{0.3}$ composite target. The prediction of the regression model is compared to the estimated ground truth.}
\label{fig:ED_AD_example_case}
\end{figure}

The magnitudes of the predicted distributions assessed through integrated quantities provide further confirmation with reference to the ground truth. Integrating once over the polar angle (energy), corresponding energy (angular) distributions are obtained. Both are presented in Figure~\ref{fig:ED_AD_example_case}. Evidently, an excellent agreement of the prediction and the ground truth is demonstrated with only minimal deviations detectable by visual inspection. This observation further manifests when integrating over the remaining energy (polar angle) coordinate and normalizing the result to the number of impinging ion projectiles. It gives the sputtering yield $Y_S$ for each species $S$ separately. Whereas the regression model predicts yields of $Y_\mathrm{Al}=0.292$, $Y_\mathrm{Ar} = 0.105$, and $Y_\mathrm{Ti} = 0.593$ for Al, Ar and Ti, respectively, the corresponding ground truth values are $Y_\mathrm{Al}=0.293$, $Y_\mathrm{Ar} = 0.102$, and $Y_\mathrm{Ti} = 0.590$. Its agreement to the second significant digit demonstrates a reliable physical description for the given sputtering scenario. 

\begin{figure}[t]
\begin{center}
\resizebox{16cm}{!}{
\includegraphics[width=16cm]{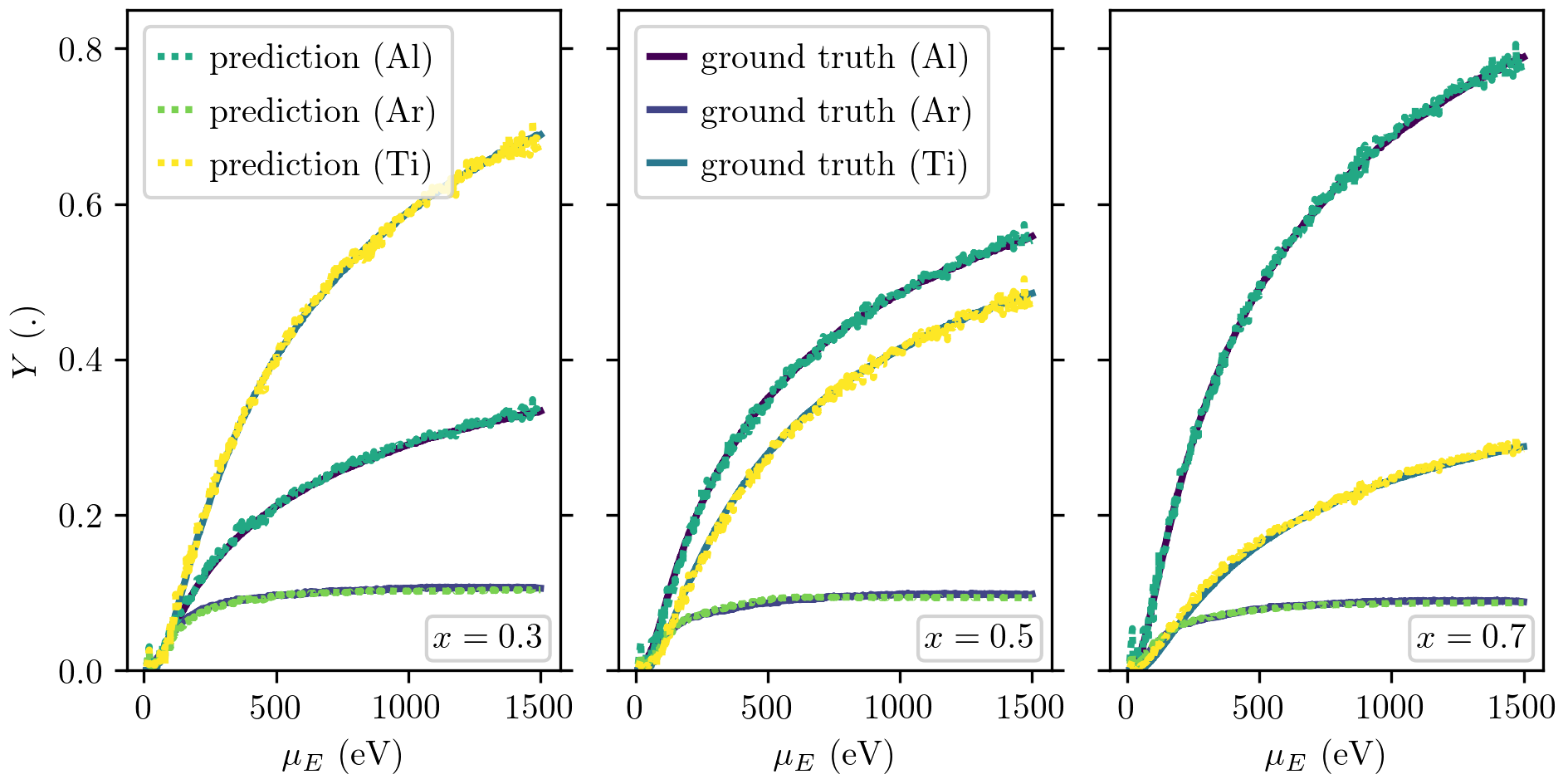}
}
\caption{Yield per sputtered species (i.e., Al, Al and Ti) as a function of the mean of any considered IED and stoichiometry.}
\label{fig:yields}
\end{center}
\end{figure}

To obtain a global trend of the model behavior, the yields $Y_S$ of all species and for all considered IEDs as well as chemical compositions are shown in Figure~\ref{fig:yields} as a function of the IED mean energy $\mu_E$, along with a comparison to the ground truth. As apparent there is an overall agreement with all curves closely overlapping. The yield is only slightly off with a maximal absolute error of 0.054 and a mean absolute error of 0.004. However, there also is a noticeable deviation for $\mu_E \approx 0$. This may be caused by a certain bias entailed in the data set (i.e., the signal to noise ratio changes varying energy). However, more probable it may be attributed to the beforehand outlined discrepancy of the latent space representations of the $\beta$-VAE and regression model for low energies. Aside from the stoichiometry distinction, the regression model maps those inputs to a latent space region (i.e., $z_2 \in [-3.76, -3.27]$) which has not been seen by the primary decoder during its training as part of the $\beta$-VAE.

Moreover, the regression model has to adapt different numerical values for all species at high energy, while at low energy the yields all approach zero. The shape of the distribution also changes, which needs to be captured regardless (cf. the surface binding energies are correlated with the EAD maxima). For the steady state situation, this may addressed by transforming the data set \textit{a priori} and considering yields normalized to the relative concentration $x_S$ of that species in the solid, which eases the regression problem. Due to flux balance constraints, $Y_S/x_S$ has to equal out in this case. However, the corresponding values $x_S$ transiently change in the dynamic case, rendering this approach infeasible. Consequently, it has not been pursued in the present work to not limit the analysis to such specific cases.

\begin{figure}[t]
\begin{center}
\resizebox{8cm}{!}{
\includegraphics[width=16cm]{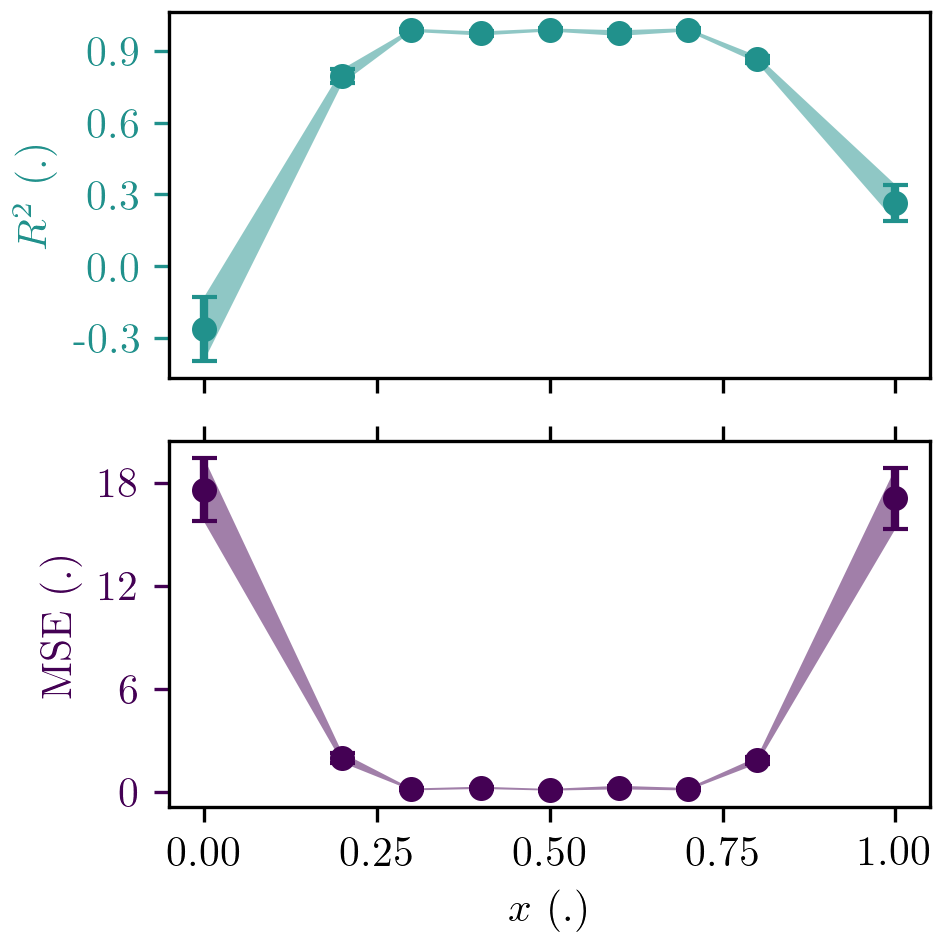}
}
\caption{The loss metrics are displayed as a function of the stoichiometry $x$ of the Ti$_{1-x}$Al$_x$ composite target. The MSE and $R^2$ error bars equal $\mu_\mathrm{MSE}\pm\sigma_\mathrm{MSE}$ and $\mu_{R^2}\pm\sigma_{R^2}$, respectively.}
\label{fig:EvaluateStoichiometry_regression}
\end{center}
\end{figure}

Finally, the capabilities of the regression model to successfully apply interpolation (and possibly extrapolation) in previously not trained cases is investigated. This is studied by taking additional Ti$_{1-x}$Al$_x$ composite target stoichiometries $x \in \{ 0.0, 0.2, 0.4, 0.6, 0.8, 1.0 \}$ for the ground truth data set into account. The network has been trained on $x \in \{ 0.3, 0.5, 0.7 \}$ only. The corresponding MSE$_\mathrm{EAD}$ and $R^2_\mathrm{EAD}$ metrics are summarized in Figure~\ref{fig:EvaluateStoichiometry_regression}. A clear trend is visible for both metrics: i) The regression ANN is most suitable to successfully perform the interpolation task $x \in \{ 0.4, 0.6 \}$ with a consistent low MSE and nearly unnoticeable influence on $R^2$. ii) In contrast, the proposed model has only a very limited capability to extrapolate into parameter regions outside the training range $x \in \{ 0.0, 0.2, 0.8, 1.0 \}$. While it gives practically useless results far away from the trained domain, it gives better but less certain results close to the training domain. This is attributed to the models ability to generalize, paired with the $\beta$-VAE approach, effectively statistically sampling also outside the original parameter space (given the respective latent space standard deviation $\sigma_\mathrm{L}$). Consequently, extrapolation is usually discouraged in the frame of machine learning. Notably, the MSE standard deviation $\sigma_\mathrm{MSE}$ from the ensemble of ANNs with identical HPs (indicated by the error bars in Figure~\ref{fig:EvaluateStoichiometry_regression}) provides a metric for the uncertainty of the prediction. This is, for instance, utilized in the context of active learning to initiate the calculation of additional data samples when the uncertainty is above a predefined threshold \cite{diaw_multiscale_2020}.

%%%%%%%%%%%%%%%%%%%%%%%%%%%%%%%%%%%%%%%%%%%%%%%%%%%
\section{Conclusion}
\label{sec:conclusion}

This work aims to extend the procedure proposed in \cite{kruger_machine_2019} to a wider range of surface materials, while making it more robust against data limitations (e.g., when using MD). In particular, the stoichiometry $x$ of the Ti-Al composite as the surface state is included as an additional input parameter. Most importantly, a $\beta$-VAE network structure is proposed, which remedies the limitations of the original MLP. The latter consists of approximately 4 million degrees of freedom, which potentially complicate the model to an unnecessary and unreliable extent. In contrast, the employed regression model consists of a total of 15,597 weights (0.390 \% of the MLP). Out of which 15,111 weights (0.378 \%) belong to the EAD decoder, which is trained as part of the $\beta$-VAE. Afterwards, these weights are set non-trainable and the learning progress is transferred by reusing the EAD decoder in the frame of the mapper/decoder regression model. Therein, only 486 weights (0.012 \%) have to be trained for the targeted regression task.

The convolutional $\beta$-VAE is trained to reduce the dimensionality of the EADs (30 energy bins $\times$ 20 angle bins $\times$ 3 species), to a 2 dimensional latent space. In addition to the typical encoder--decoder pair, a secondary decoder is introduced to condition and simultaneously interpret the obtained latent space in terms of the incident IED's mean energy $\mu_E^\prime$ and Ti-Al stoichiometry $x^\prime$. The utilized set of hyperparameters that minimizes the reconstruction loss while keeping the model as simple as possible is determined by a hyperparameter study. As worked out in the discussion of the latent space, the $\beta$-VAE trains the model not only at the given data points, but also in their vicinity. The utilized concept is partially similar to a conditional VAE \cite{sohn_learning_2015,doersch_tutorial_2021}. Therein, the secondary decoder is omitted, while the encoder and the decoder are both provided with the conditional input information. Simultaneously, the reconstructed output variables are reduced and subsequently reconstructed. Hence, the latent space is affected by the input variables in both models in different ways. Moreover, the conditional VAE is a regression model itself, while the here presented $\beta$-VAE is meant to reduce the EAD dimensionality. The obtained encoder/decoder networks are prevalently used to project into/from the low-dimensional latent space, while regression is achieved via a data-minimalistic mapping through a dedicated ANN component.

The assembled regression model has to be trained to map the input variables to the latent space. This network has been optimized through a corresponding hyperparameter study. A final test on the estimated ground truth yields a mean $\mu_{R^2} = 0.991$ and a standard deviation $\sigma_{R^2} = 0.004$ for the coefficient of determination over a 10-fold cross validation. The corresponding values for the complete data set including the unknown interpolation stoichiometries $x=\{0.4, 0.6\}$ are $\mu_{R^2} = 0.987$ and  $\sigma_{R^2} = 0.004$. This excellent agreement of the EADs over the complete data set as well as for the exemplified case of the integrated energy distribution, angular distribution and yield supports that the model is able to describe the sputtering process for a wide range of IED as well as Ti-Al composites. Notably, it is in general not feasible to estimate the ground truth (e.g., when using MD simulations or experimental data). The ground truth data is not required for the training procedure, but is useful for assessing its performance.

The mean value and the standard deviation of the time required by the regression network to predict a single EAD are found to be 0.310 ms and 0.002 ms, respectively, on a laptop computer equipped with an Intel i7-10510U CPU. Thus, it can be readily applied in the frame of gas-phase simulations as an interface model, as outlined elsewhere \cite{kruger_machine_2019}. Application to more sophisticated surface states described by more than only its stoichiometry, as well as data-limited reactive molecular dynamics simulations potentially causing an even more challenging statistical representation have to be addressed in a future work. This is suggested to establish a plasma-surface interface model for complex materials and system dynamics. Additionally, the options to include inherent physical constraints (e.g., flux balance in the transient situation with varying stoichiometry) are suggested for exploration.

%%%%%%%%%%%%%%%%%%%%%%%%%%%%%%%%%%%%%%%%%%%%%%%%%%%
\section*{Acknowledgement}
The authors sincerely thank Professor Dr.-Ing.\ Thomas Mussenbrock from Ruhr University Bochum for his support. The authors thank Professor Dr.\ Wolfhard Möller from Institute of Ion Beam Physics and Materials Research, Helmholtz-Zentrum Dresden-Rossendorf (HZDR) for permission to use the TRIDYN simulation software. Funded by the Deutsche Forschungsgemeinschaft (DFG, German Research Foundation) -- Project-ID 138690629 -- TRR 87 and -- Project-ID 434434223 -- SFB 1461.

%%%%%%%%%%%%%%%%%%%%%%%%%%%%%%%%%%%%%%%%%%%%%%%%%%%

\section*{ORCID iDs}
\noindent
T. Gergs: \url{https://orcid.org/0000-0001-5041-2941}\\
B. Borislavov: \url{https://orcid.org/0000-0001-7753-4156}\\
J. Trieschmann: \url{https://orcid.org/0000-0001-9136-8019}

%%%%%%%%%%%%%%%%%%%%%%%%%%%%%%%%%%%%%%%%%%%%%%%%%%%

\appendix
\section*{Appendix}

\begin{table}[b]
\caption{General TRIDYN parameters.}
\label{table:global tridyn_parameters}
\begin{center}
\begin{tabular}{ l   c   c }
\hline
parameter & symbol & value\\
\hline
no. of projectiles & $N_\text{sp}$ & $10^4$ and $10^6$ \\
total fluence & $\Phi_\text{tot}$ & $1 \text{ \AA}^{-2}$ \\
angle of irradiation & $\vartheta_0$  & 0 \\
max. depth & $x_\text{max}$ &  600 \text{\AA} \\
depth interval & $\Delta x $ & 3 \text{\AA} \\
Ar max. atomic fraction & $x_\text{Ar}$ & 0.1 \\
\hline
\end{tabular}
\end{center}
\end{table}

\begin{table}[t]
\caption{Element specific TRIDYN parameters. $\varepsilon_{s,t}$ denotes the surface binding energy matrix elements, where index combinations $ s, t \,\epsilon \, \{ \text{Ar, Al, Ti} \}$ iterate over the individual contents of each target component, $ Qu_s $ is the respective initial atomic fraction, and $n_s$ the atomic density.}
\label{table:element specific tridyn_parameters}
\begin{center}

\begin{tabular}{ l   c    c   c   c   c}
\hline
element  & $\, \varepsilon_{s,\text{Ar}}$ (eV) &\,  $\varepsilon_{s,\text{Al}}$ (eV) &\,  $\varepsilon_{s,\text{Ti}}$ (eV) & \, $ Qu_s $ \,&\,  $n_s$ ($\text{\AA}^{-3}$) \\
\hline
Ar & 0 & 0    & 0		& 0     & 0.0249 \\
Al & 0 & 3.36 & 4.12	& $x$   & 0.0602 \\
Ti & 0 & 4.12 & 4.89	& $1-x$ & 0.0558 \\
\hline

\end{tabular}
\end{center}
\end{table}

\begin{figure*}[h]
\includegraphics[width=16cm]{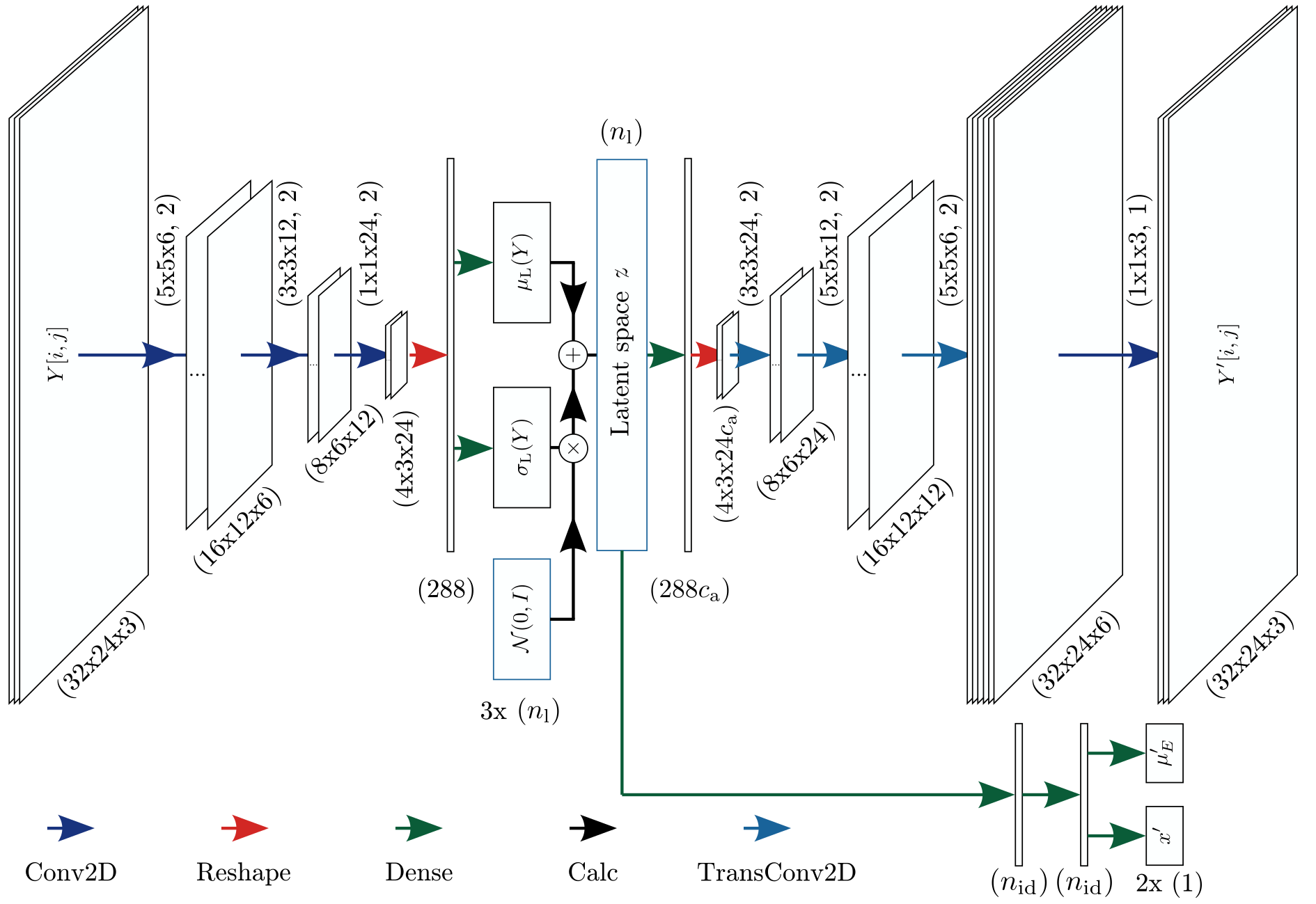}
\caption{Schematic of the $\beta$-VAE network structure. The shape of the data output by the intermediate layers is indicated at the data pictographs. The artificial neural network operations are indicated by colored arrows, whereas the convolutional operations are detailed above the arrows by the corresponding kernel size $k$, number of filters $f$, and stride $s$, for example $(k \times k \times f, s)$.}
\label{fig:VAE_network_detail}
\end{figure*}

\begin{figure*}[h]
\includegraphics[width=16cm]{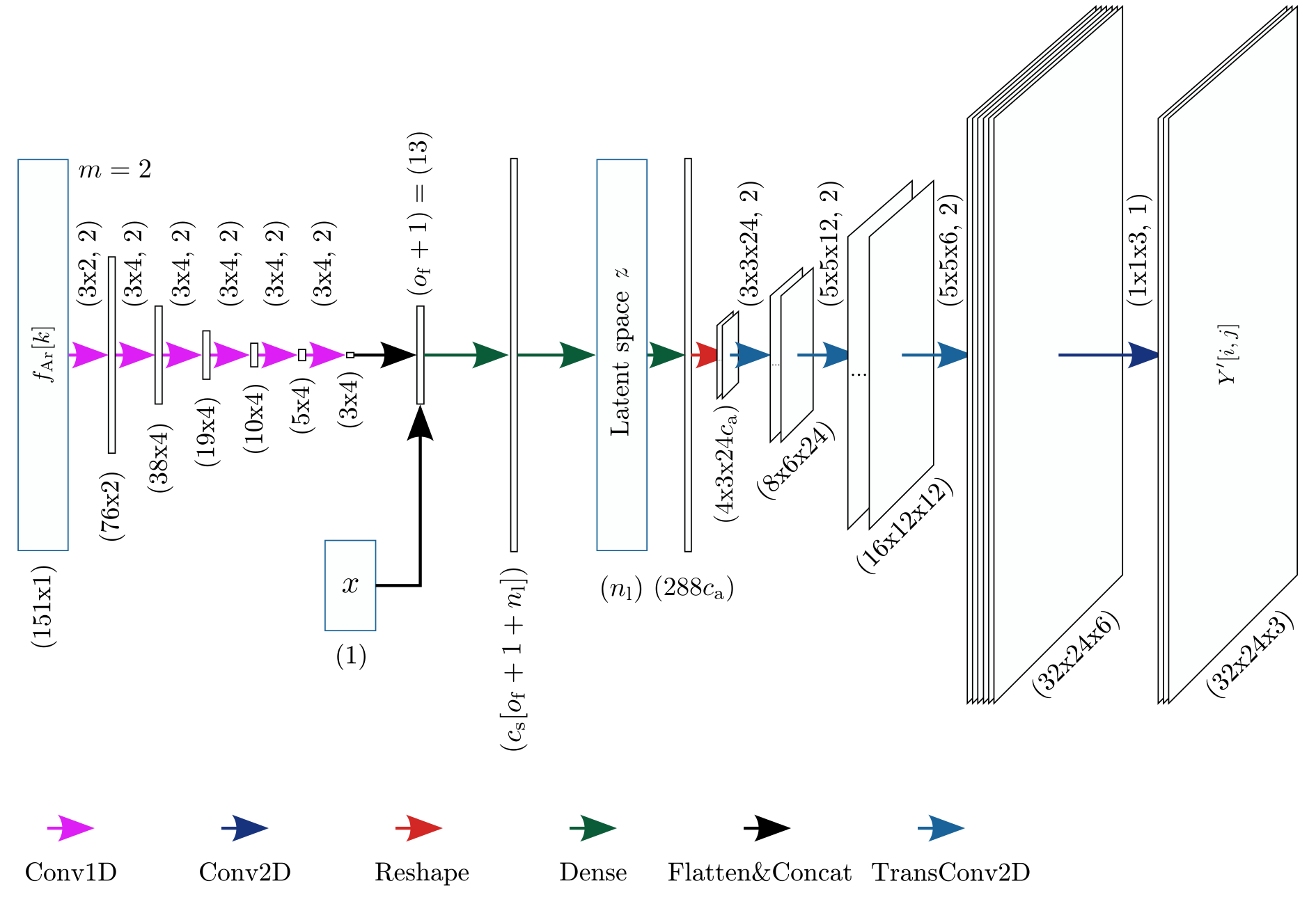}
\caption{Schematic of the regression network structure. The shape of the data output by the intermediate layers is indicated at the data pictographs. The artificial neural network operations are indicated by colored arrows, whereas the convolutional operations are detailed above the arrows by the corresponding kernel size $k$, number of filters $f$, and stride $s$, for example $(k \times k \times f, s)$.}
\label{fig:regression_network_detail}
\end{figure*}

%%%%%%%%%%%%%%%%%%%%%%%%%%%%%%%%%%%%%%%%%%%%%%%%%%%
\clearpage
\bibliography{references}

%merlin.mbs apsrev4-1.bst 2010-07-25 4.21a (PWD, AO, DPC) hacked
%Control: key (0)
%Control: author (72) initials jnrlst
%Control: editor formatted (1) identically to author
%Control: production of article title (-1) disabled
%Control: page (0) single
%Control: year (1) truncated
%Control: production of eprint (0) enabled
\begin{thebibliography}{39}%
\makeatletter
\providecommand \@ifxundefined [1]{%
 \@ifx{#1\undefined}
}%
\providecommand \@ifnum [1]{%
 \ifnum #1\expandafter \@firstoftwo
 \else \expandafter \@secondoftwo
 \fi
}%
\providecommand \@ifx [1]{%
 \ifx #1\expandafter \@firstoftwo
 \else \expandafter \@secondoftwo
 \fi
}%
\providecommand \natexlab [1]{#1}%
\providecommand \enquote  [1]{``#1''}%
\providecommand \bibnamefont  [1]{#1}%
\providecommand \bibfnamefont [1]{#1}%
\providecommand \citenamefont [1]{#1}%
\providecommand \href@noop [0]{\@secondoftwo}%
\providecommand \href [0]{\begingroup \@sanitize@url \@href}%
\providecommand \@href[1]{\@@startlink{#1}\@@href}%
\providecommand \@@href[1]{\endgroup#1\@@endlink}%
\providecommand \@sanitize@url [0]{\catcode `\\12\catcode `\$12\catcode
  `\&12\catcode `\#12\catcode `\^12\catcode `\_12\catcode `\%12\relax}%
\providecommand \@@startlink[1]{}%
\providecommand \@@endlink[0]{}%
\providecommand \url  [0]{\begingroup\@sanitize@url \@url }%
\providecommand \@url [1]{\endgroup\@href {#1}{\urlprefix }}%
\providecommand \urlprefix  [0]{URL }%
\providecommand \Eprint [0]{\href }%
\providecommand \doibase [0]{http://dx.doi.org/}%
\providecommand \selectlanguage [0]{\@gobble}%
\providecommand \bibinfo  [0]{\@secondoftwo}%
\providecommand \bibfield  [0]{\@secondoftwo}%
\providecommand \translation [1]{[#1]}%
\providecommand \BibitemOpen [0]{}%
\providecommand \bibitemStop [0]{}%
\providecommand \bibitemNoStop [0]{.\EOS\space}%
\providecommand \EOS [0]{\spacefactor3000\relax}%
\providecommand \BibitemShut  [1]{\csname bibitem#1\endcsname}%
\let\auto@bib@innerbib\@empty
%</preamble>
\bibitem [{\citenamefont {Krüger}\ \emph {et~al.}(2019)\citenamefont
  {Krüger}, \citenamefont {Gergs},\ and\ \citenamefont
  {Trieschmann}}]{kruger_machine_2019}%
  \BibitemOpen
  \bibfield  {author} {\bibinfo {author} {\bibfnamefont {F.}~\bibnamefont
  {Krüger}}, \bibinfo {author} {\bibfnamefont {T.}~\bibnamefont {Gergs}}, \
  and\ \bibinfo {author} {\bibfnamefont {J.}~\bibnamefont {Trieschmann}},\
  }\href {\doibase 10.1088/1361-6595/ab0246} {\bibfield  {journal} {\bibinfo
  {journal} {Plasma Sources Science and Technology}\ }\textbf {\bibinfo
  {volume} {28}},\ \bibinfo {pages} {035002} (\bibinfo {year}
  {2019})}\BibitemShut {NoStop}%
\bibitem [{\citenamefont {Bird}(1994)}]{bird_molecular_1994}%
  \BibitemOpen
  \bibfield  {author} {\bibinfo {author} {\bibfnamefont {G.~A.}\ \bibnamefont
  {Bird}},\ }\href@noop {} {{\selectlanguage {English}\emph {\bibinfo {title}
  {Molecular {Gas} {Dynamics} and the {Direct} {Simulation} of {Gas}
  {Flows}}}}}\ (\bibinfo  {publisher} {Oxford University Press},\ \bibinfo
  {address} {New York, USA},\ \bibinfo {year} {1994})\BibitemShut {NoStop}%
\bibitem [{\citenamefont {Lieberman}\ and\ \citenamefont
  {Lichtenberg}(2005)}]{lieberman_principles_2005}%
  \BibitemOpen
  \bibfield  {author} {\bibinfo {author} {\bibfnamefont {M.~A.}\ \bibnamefont
  {Lieberman}}\ and\ \bibinfo {author} {\bibfnamefont {A.~J.}\ \bibnamefont
  {Lichtenberg}},\ }\href {https://dx.doi.org/10.1002/0471724254} {\emph
  {\bibinfo {title} {Principles of {Plasma} {Discharges} and {Materials}
  {Processing}}}},\ \bibinfo {edition} {2nd}\ ed.\ (\bibinfo  {publisher}
  {Wiley},\ \bibinfo {address} {Hoboken, USA},\ \bibinfo {year}
  {2005})\BibitemShut {NoStop}%
\bibitem [{\citenamefont {Callister}\ and\ \citenamefont
  {Rethwisch}(2013)}]{callister_materials_2013}%
  \BibitemOpen
  \bibfield  {author} {\bibinfo {author} {\bibfnamefont {W.~D.~J.}\
  \bibnamefont {Callister}}\ and\ \bibinfo {author} {\bibfnamefont {D.~G.}\
  \bibnamefont {Rethwisch}},\ }\href@noop {} {\emph {\bibinfo {title}
  {Materials {Science} and {Engineering}: {An} {Introduction}}}},\ \bibinfo
  {edition} {9th}\ ed.\ (\bibinfo  {publisher} {Wiley},\ \bibinfo {address}
  {Hoboken, USA},\ \bibinfo {year} {2013})\BibitemShut {NoStop}%
\bibitem [{\citenamefont {Biersack}\ and\ \citenamefont
  {Haggmark}(1980)}]{biersack_monte_1980}%
  \BibitemOpen
  \bibfield  {author} {\bibinfo {author} {\bibfnamefont {J.~P.}\ \bibnamefont
  {Biersack}}\ and\ \bibinfo {author} {\bibfnamefont {L.~G.}\ \bibnamefont
  {Haggmark}},\ }\href {\doibase 10.1016/0029-554X(80)90440-1} {\bibfield
  {journal} {\bibinfo  {journal} {Nuclear Instruments and Methods}\ }\textbf
  {\bibinfo {volume} {174}},\ \bibinfo {pages} {257} (\bibinfo {year}
  {1980})}\BibitemShut {NoStop}%
\bibitem [{\citenamefont {Eckstein}\ and\ \citenamefont
  {Biersack}(1984)}]{eckstein_sputtering_1984}%
  \BibitemOpen
  \bibfield  {author} {\bibinfo {author} {\bibfnamefont {W.}~\bibnamefont
  {Eckstein}}\ and\ \bibinfo {author} {\bibfnamefont {J.}~\bibnamefont
  {Biersack}},\ }\href {\doibase 10.1016/0168-583X(84)90264-7} {\bibfield
  {journal} {\bibinfo  {journal} {Nuclear Instruments and Methods in Physics
  Research Section B: Beam Interactions with Materials and Atoms}\ }\textbf
  {\bibinfo {volume} {2}},\ \bibinfo {pages} {550} (\bibinfo {year}
  {1984})}\BibitemShut {NoStop}%
\bibitem [{\citenamefont {Möller}\ and\ \citenamefont
  {Eckstein}(1984)}]{moller_tridyn_1984}%
  \BibitemOpen
  \bibfield  {author} {\bibinfo {author} {\bibfnamefont {W.}~\bibnamefont
  {Möller}}\ and\ \bibinfo {author} {\bibfnamefont {W.}~\bibnamefont
  {Eckstein}},\ }\href {\doibase 10.1016/0168-583X(84)90321-5} {\bibfield
  {journal} {\bibinfo  {journal} {Nuclear Instruments and Methods in Physics
  Research Section B: Beam Interactions with Materials and Atoms}\ }\textbf
  {\bibinfo {volume} {2}},\ \bibinfo {pages} {814} (\bibinfo {year}
  {1984})}\BibitemShut {NoStop}%
\bibitem [{\citenamefont {Voter}(2007)}]{voter_introduction_2007}%
  \BibitemOpen
  \bibfield  {author} {\bibinfo {author} {\bibfnamefont {A.~F.}\ \bibnamefont
  {Voter}},\ }in\ \href {https://dx.doi.org/10.1007/978-1-4020-5295-8_1} {\emph
  {\bibinfo {booktitle} {Radiation effects in solids}}},\ \bibinfo {series and
  number} {\bibinfo {series} {{NATO} science series. {II}, {Mathematics},
  physics and chemistry}\ No.\ \bibinfo {number} {v. 235}}\ (\bibinfo
  {publisher} {Springer},\ \bibinfo {address} {Dordrecht, The Netherlands},\
  \bibinfo {year} {2007})\BibitemShut {NoStop}%
\bibitem [{\citenamefont {Graves}\ and\ \citenamefont
  {Brault}(2009)}]{graves_molecular_2009}%
  \BibitemOpen
  \bibfield  {author} {\bibinfo {author} {\bibfnamefont {D.~B.}\ \bibnamefont
  {Graves}}\ and\ \bibinfo {author} {\bibfnamefont {P.}~\bibnamefont
  {Brault}},\ }\href {\doibase 10.1088/0022-3727/42/19/194011} {\bibfield
  {journal} {\bibinfo  {journal} {Journal of Physics D: Applied Physics}\
  }\textbf {\bibinfo {volume} {42}},\ \bibinfo {pages} {194011} (\bibinfo
  {year} {2009})}\BibitemShut {NoStop}%
\bibitem [{\citenamefont {Neyts}\ and\ \citenamefont
  {Brault}(2017)}]{neyts_molecular_2017}%
  \BibitemOpen
  \bibfield  {author} {\bibinfo {author} {\bibfnamefont {E.~C.}\ \bibnamefont
  {Neyts}}\ and\ \bibinfo {author} {\bibfnamefont {P.}~\bibnamefont {Brault}},\
  }\href {\doibase 10.1002/ppap.201600145} {\bibfield  {journal} {\bibinfo
  {journal} {Plasma Processes and Polymers}\ }\textbf {\bibinfo {volume}
  {14}},\ \bibinfo {pages} {1600145} (\bibinfo {year} {2017})}\BibitemShut
  {NoStop}%
\bibitem [{\citenamefont {Birdsall}\ and\ \citenamefont
  {Langdon}(1991)}]{birdsall_plasma_1991}%
  \BibitemOpen
  \bibfield  {author} {\bibinfo {author} {\bibfnamefont {C.~K.}\ \bibnamefont
  {Birdsall}}\ and\ \bibinfo {author} {\bibfnamefont {A.~B.}\ \bibnamefont
  {Langdon}},\ }\href@noop {} {\emph {\bibinfo {title} {Plasma {Physics} via
  {Computer} {Simulations}}}}\ (\bibinfo  {publisher} {IOP Publishing},\
  \bibinfo {address} {Bristol, UK},\ \bibinfo {year} {1991})\BibitemShut
  {NoStop}%
\bibitem [{\citenamefont {Dijk}\ \emph {et~al.}(2009)\citenamefont {Dijk},
  \citenamefont {Kroesen},\ and\ \citenamefont {Bogaerts}}]{dijk_plasma_2009}%
  \BibitemOpen
  \bibfield  {author} {\bibinfo {author} {\bibfnamefont {J.~v.}\ \bibnamefont
  {Dijk}}, \bibinfo {author} {\bibfnamefont {G.~M.~W.}\ \bibnamefont
  {Kroesen}}, \ and\ \bibinfo {author} {\bibfnamefont {A.}~\bibnamefont
  {Bogaerts}},\ }\href {\doibase 10.1088/0022-3727/42/19/190301} {\bibfield
  {journal} {\bibinfo  {journal} {Journal of Physics D: Applied Physics}\
  }\textbf {\bibinfo {volume} {42}},\ \bibinfo {pages} {190301} (\bibinfo
  {year} {2009})}\BibitemShut {NoStop}%
\bibitem [{\citenamefont {Serikov}\ \emph {et~al.}(1999)\citenamefont
  {Serikov}, \citenamefont {Kawamoto},\ and\ \citenamefont
  {Nanbu}}]{serikov_particle--cell_1999}%
  \BibitemOpen
  \bibfield  {author} {\bibinfo {author} {\bibfnamefont {V.}~\bibnamefont
  {Serikov}}, \bibinfo {author} {\bibfnamefont {S.}~\bibnamefont {Kawamoto}}, \
  and\ \bibinfo {author} {\bibfnamefont {K.}~\bibnamefont {Nanbu}},\ }\href
  {\doibase 10.1109/27.799817} {\bibfield  {journal} {\bibinfo  {journal} {IEEE
  Transactions on Plasma Science}\ }\textbf {\bibinfo {volume} {27}},\ \bibinfo
  {pages} {1389} (\bibinfo {year} {1999})}\BibitemShut {NoStop}%
\bibitem [{\citenamefont {Somekh}(1984)}]{somekh_thermalization_1984}%
  \BibitemOpen
  \bibfield  {author} {\bibinfo {author} {\bibfnamefont {R.~E.}\ \bibnamefont
  {Somekh}},\ }\href {\doibase 10.1116/1.572396} {\bibfield  {journal}
  {\bibinfo  {journal} {Journal of Vacuum Science \& Technology A}\ }\textbf
  {\bibinfo {volume} {2}},\ \bibinfo {pages} {1285} (\bibinfo {year}
  {1984})}\BibitemShut {NoStop}%
\bibitem [{\citenamefont {Turner}\ \emph {et~al.}(1989)\citenamefont {Turner},
  \citenamefont {Falconer}, \citenamefont {James},\ and\ \citenamefont
  {McKenzie}}]{turner_monte_1989}%
  \BibitemOpen
  \bibfield  {author} {\bibinfo {author} {\bibfnamefont {G.~M.}\ \bibnamefont
  {Turner}}, \bibinfo {author} {\bibfnamefont {I.~S.}\ \bibnamefont
  {Falconer}}, \bibinfo {author} {\bibfnamefont {B.~W.}\ \bibnamefont {James}},
  \ and\ \bibinfo {author} {\bibfnamefont {D.~R.}\ \bibnamefont {McKenzie}},\
  }\href {\doibase 10.1063/1.342593} {\bibfield  {journal} {\bibinfo  {journal}
  {Journal of Applied Physics}\ }\textbf {\bibinfo {volume} {65}},\ \bibinfo
  {pages} {3671} (\bibinfo {year} {1989})}\BibitemShut {NoStop}%
\bibitem [{\citenamefont {Trieschmann}\ and\ \citenamefont
  {Mussenbrock}(2015)}]{trieschmann_transport_2015}%
  \BibitemOpen
  \bibfield  {author} {\bibinfo {author} {\bibfnamefont {J.}~\bibnamefont
  {Trieschmann}}\ and\ \bibinfo {author} {\bibfnamefont {T.}~\bibnamefont
  {Mussenbrock}},\ }\href {\doibase 10.1063/1.4926878} {\bibfield  {journal}
  {\bibinfo  {journal} {Journal of Applied Physics}\ }\textbf {\bibinfo
  {volume} {118}},\ \bibinfo {pages} {033302} (\bibinfo {year}
  {2015})}\BibitemShut {NoStop}%
\bibitem [{\citenamefont {Neyts}\ \emph {et~al.}(2010)\citenamefont {Neyts},
  \citenamefont {Shibuta}, \citenamefont {van Duin},\ and\ \citenamefont
  {Bogaerts}}]{neyts_catalyzed_2010}%
  \BibitemOpen
  \bibfield  {author} {\bibinfo {author} {\bibfnamefont {E.~C.}\ \bibnamefont
  {Neyts}}, \bibinfo {author} {\bibfnamefont {Y.}~\bibnamefont {Shibuta}},
  \bibinfo {author} {\bibfnamefont {A.~C.~T.}\ \bibnamefont {van Duin}}, \ and\
  \bibinfo {author} {\bibfnamefont {A.}~\bibnamefont {Bogaerts}},\ }\href
  {\doibase 10.1021/nn102095y} {\bibfield  {journal} {\bibinfo  {journal} {ACS
  Nano}\ }\textbf {\bibinfo {volume} {4}},\ \bibinfo {pages} {6665} (\bibinfo
  {year} {2010})}\BibitemShut {NoStop}%
\bibitem [{\citenamefont {Neyts}\ and\ \citenamefont
  {Bogaerts}(2013)}]{neyts_combining_2012}%
  \BibitemOpen
  \bibfield  {author} {\bibinfo {author} {\bibfnamefont {E.~C.}\ \bibnamefont
  {Neyts}}\ and\ \bibinfo {author} {\bibfnamefont {A.}~\bibnamefont
  {Bogaerts}},\ }\href {\doibase 10.1007/s00214-012-1320-x} {\bibfield
  {journal} {\bibinfo  {journal} {Theoretical Chemistry Accounts}\ }\textbf
  {\bibinfo {volume} {132}},\ \bibinfo {pages} {1320} (\bibinfo {year}
  {2013})}\BibitemShut {NoStop}%
\bibitem [{\citenamefont {Tonneau}\ \emph {et~al.}(2018)\citenamefont
  {Tonneau}, \citenamefont {Moskovkin}, \citenamefont {Pflug},\ and\
  \citenamefont {Lucas}}]{tonneau_tioxdeposited_2018}%
  \BibitemOpen
  \bibfield  {author} {\bibinfo {author} {\bibfnamefont {R.}~\bibnamefont
  {Tonneau}}, \bibinfo {author} {\bibfnamefont {P.}~\bibnamefont {Moskovkin}},
  \bibinfo {author} {\bibfnamefont {A.}~\bibnamefont {Pflug}}, \ and\ \bibinfo
  {author} {\bibfnamefont {S.}~\bibnamefont {Lucas}},\ }\href {\doibase
  10.1088/1361-6463/aabb72} {\bibfield  {journal} {\bibinfo  {journal} {Journal
  of Physics D: Applied Physics}\ }\textbf {\bibinfo {volume} {51}},\ \bibinfo
  {pages} {195202} (\bibinfo {year} {2018})}\BibitemShut {NoStop}%
\bibitem [{\citenamefont {Thompson}(1968)}]{thompson_ii_1968}%
  \BibitemOpen
  \bibfield  {author} {\bibinfo {author} {\bibfnamefont {M.~W.}\ \bibnamefont
  {Thompson}},\ }\href {\doibase 10.1080/14786436808227358} {\bibfield
  {journal} {\bibinfo  {journal} {The Philosophical Magazine: A Journal of
  Theoretical Experimental and Applied Physics}\ }\textbf {\bibinfo {volume}
  {18}},\ \bibinfo {pages} {377} (\bibinfo {year} {1968})}\BibitemShut
  {NoStop}%
\bibitem [{\citenamefont {Sigmund}(1969{\natexlab{a}})}]{sigmund_theory_1969}%
  \BibitemOpen
  \bibfield  {author} {\bibinfo {author} {\bibfnamefont {P.}~\bibnamefont
  {Sigmund}},\ }\href {\doibase 10.1103/PhysRev.187.768} {\bibfield  {journal}
  {\bibinfo  {journal} {Physical Review}\ }\textbf {\bibinfo {volume} {187}},\
  \bibinfo {pages} {768} (\bibinfo {year} {1969}{\natexlab{a}})}\BibitemShut
  {NoStop}%
\bibitem [{\citenamefont
  {Sigmund}(1969{\natexlab{b}})}]{sigmund_theory_1969-1}%
  \BibitemOpen
  \bibfield  {author} {\bibinfo {author} {\bibfnamefont {P.}~\bibnamefont
  {Sigmund}},\ }\href {\doibase 10.1103/PhysRev.184.383} {\bibfield  {journal}
  {\bibinfo  {journal} {Physical Review}\ }\textbf {\bibinfo {volume} {184}},\
  \bibinfo {pages} {383} (\bibinfo {year} {1969}{\natexlab{b}})}\BibitemShut
  {NoStop}%
\bibitem [{\citenamefont {Berg}\ and\ \citenamefont
  {Nyberg}(2005)}]{berg_fundamental_2005}%
  \BibitemOpen
  \bibfield  {author} {\bibinfo {author} {\bibfnamefont {S.}~\bibnamefont
  {Berg}}\ and\ \bibinfo {author} {\bibfnamefont {T.}~\bibnamefont {Nyberg}},\
  }\href {\doibase 10.1016/j.tsf.2004.10.051} {\bibfield  {journal} {\bibinfo
  {journal} {Thin Solid Films}\ }\textbf {\bibinfo {volume} {476}},\ \bibinfo
  {pages} {215} (\bibinfo {year} {2005})}\BibitemShut {NoStop}%
\bibitem [{\citenamefont {Depla}\ \emph {et~al.}(2008)\citenamefont {Depla},
  \citenamefont {Mahieu}, \citenamefont {Hull}, \citenamefont {Osgood},
  \citenamefont {Parisi},\ and\ \citenamefont
  {Warlimont}}]{depla_reactive_2008}%
  \BibitemOpen
  \bibinfo {editor} {\bibfnamefont {D.}~\bibnamefont {Depla}}, \bibinfo
  {editor} {\bibfnamefont {S.}~\bibnamefont {Mahieu}}, \bibinfo {editor}
  {\bibfnamefont {R.}~\bibnamefont {Hull}}, \bibinfo {editor} {\bibfnamefont
  {R.~M.}\ \bibnamefont {Osgood}}, \bibinfo {editor} {\bibfnamefont
  {J.}~\bibnamefont {Parisi}}, \ and\ \bibinfo {editor} {\bibfnamefont
  {H.}~\bibnamefont {Warlimont}},\ eds.,\ \href
  {https://dx.doi.org/10.1007/978-3-540-76664-3} {\emph {\bibinfo {title}
  {Reactive {Sputter} {Deposition}}}},\ \bibinfo {series} {Springer {Series} in
  {Materials} {Science}}, Vol.\ \bibinfo {volume} {109}\ (\bibinfo  {publisher}
  {Springer},\ \bibinfo {address} {Berlin, Germany},\ \bibinfo {year}
  {2008})\BibitemShut {NoStop}%
\bibitem [{\citenamefont {Diaw}\ \emph {et~al.}(2020)\citenamefont {Diaw},
  \citenamefont {Barros}, \citenamefont {Haack}, \citenamefont {Junghans},
  \citenamefont {Keenan}, \citenamefont {Li}, \citenamefont {Livescu},
  \citenamefont {Lubbers}, \citenamefont {McKerns}, \citenamefont {Pavel},
  \citenamefont {Rosenberger}, \citenamefont {Sagert},\ and\ \citenamefont
  {Germann}}]{diaw_multiscale_2020}%
  \BibitemOpen
  \bibfield  {author} {\bibinfo {author} {\bibfnamefont {A.}~\bibnamefont
  {Diaw}}, \bibinfo {author} {\bibfnamefont {K.}~\bibnamefont {Barros}},
  \bibinfo {author} {\bibfnamefont {J.}~\bibnamefont {Haack}}, \bibinfo
  {author} {\bibfnamefont {C.}~\bibnamefont {Junghans}}, \bibinfo {author}
  {\bibfnamefont {B.}~\bibnamefont {Keenan}}, \bibinfo {author} {\bibfnamefont
  {Y.~W.}\ \bibnamefont {Li}}, \bibinfo {author} {\bibfnamefont
  {D.}~\bibnamefont {Livescu}}, \bibinfo {author} {\bibfnamefont
  {N.}~\bibnamefont {Lubbers}}, \bibinfo {author} {\bibfnamefont
  {M.}~\bibnamefont {McKerns}}, \bibinfo {author} {\bibfnamefont {R.~S.}\
  \bibnamefont {Pavel}}, \bibinfo {author} {\bibfnamefont {D.}~\bibnamefont
  {Rosenberger}}, \bibinfo {author} {\bibfnamefont {I.}~\bibnamefont {Sagert}},
  \ and\ \bibinfo {author} {\bibfnamefont {T.~C.}\ \bibnamefont {Germann}},\
  }\href {\doibase 10.1103/PhysRevE.102.023310} {\bibfield  {journal} {\bibinfo
   {journal} {Physical Review E}\ }\textbf {\bibinfo {volume} {102}},\ \bibinfo
  {pages} {023310} (\bibinfo {year} {2020})}\BibitemShut {NoStop}%
\bibitem [{\citenamefont {Ulissi}\ \emph {et~al.}(2017)\citenamefont {Ulissi},
  \citenamefont {Medford}, \citenamefont {Bligaard},\ and\ \citenamefont
  {Nørskov}}]{ulissi_address_2017}%
  \BibitemOpen
  \bibfield  {author} {\bibinfo {author} {\bibfnamefont {Z.~W.}\ \bibnamefont
  {Ulissi}}, \bibinfo {author} {\bibfnamefont {A.~J.}\ \bibnamefont {Medford}},
  \bibinfo {author} {\bibfnamefont {T.}~\bibnamefont {Bligaard}}, \ and\
  \bibinfo {author} {\bibfnamefont {J.~K.}\ \bibnamefont {Nørskov}},\ }\href
  {\doibase 10.1038/ncomms14621} {\bibfield  {journal} {\bibinfo  {journal}
  {Nature Communications}\ }\textbf {\bibinfo {volume} {8}},\ \bibinfo {pages}
  {14621} (\bibinfo {year} {2017})}\BibitemShut {NoStop}%
\bibitem [{\citenamefont {Kino}\ \emph {et~al.}(2021)\citenamefont {Kino},
  \citenamefont {Ikuse}, \citenamefont {Dam},\ and\ \citenamefont
  {Hamaguchi}}]{kino_characterization_2021}%
  \BibitemOpen
  \bibfield  {author} {\bibinfo {author} {\bibfnamefont {H.}~\bibnamefont
  {Kino}}, \bibinfo {author} {\bibfnamefont {K.}~\bibnamefont {Ikuse}},
  \bibinfo {author} {\bibfnamefont {H.-C.}\ \bibnamefont {Dam}}, \ and\
  \bibinfo {author} {\bibfnamefont {S.}~\bibnamefont {Hamaguchi}},\ }\href
  {\doibase 10.1063/5.0006816} {\bibfield  {journal} {\bibinfo  {journal}
  {Physics of Plasmas}\ }\textbf {\bibinfo {volume} {28}},\ \bibinfo {pages}
  {013504} (\bibinfo {year} {2021})}\BibitemShut {NoStop}%
\bibitem [{\citenamefont {Kingma}\ and\ \citenamefont
  {Welling}(2013)}]{kingma_auto-encoding_2013}%
  \BibitemOpen
  \bibfield  {author} {\bibinfo {author} {\bibfnamefont {D.~P.}\ \bibnamefont
  {Kingma}}\ and\ \bibinfo {author} {\bibfnamefont {M.}~\bibnamefont
  {Welling}},\ }in\ \href {https://openreview.net/forum?id=33X9fd2-9FyZd}
  {{\selectlanguage {en}\emph {\bibinfo {booktitle} {Proceedings of the
  International Conference on Learning Representations}}}}\ (\bibinfo {address}
  {Scottsdale, USA},\ \bibinfo {year} {2013})\BibitemShut {NoStop}%
\bibitem [{\citenamefont {Rezende}\ \emph {et~al.}(2014)\citenamefont
  {Rezende}, \citenamefont {Mohamed},\ and\ \citenamefont
  {Wierstra}}]{rezende_stochastic_2014}%
  \BibitemOpen
  \bibfield  {author} {\bibinfo {author} {\bibfnamefont {D.~J.}\ \bibnamefont
  {Rezende}}, \bibinfo {author} {\bibfnamefont {S.}~\bibnamefont {Mohamed}}, \
  and\ \bibinfo {author} {\bibfnamefont {D.}~\bibnamefont {Wierstra}},\ }in\
  \href {https://proceedings.mlr.press/v32/rezende14.html} {{\selectlanguage
  {en}\emph {\bibinfo {booktitle} {Proceedings of the 31st International
  Conference on Machine Learning}}}},\ Vol.~\bibinfo {volume} {32}\ (\bibinfo
  {publisher} {PMLR},\ \bibinfo {address} {Beijing, CN},\ \bibinfo {year}
  {2014})\ pp.\ \bibinfo {pages} {1278--1286}\BibitemShut {NoStop}%
\bibitem [{\citenamefont {Higgins}\ \emph {et~al.}(2016)\citenamefont
  {Higgins}, \citenamefont {Matthey}, \citenamefont {Pal}, \citenamefont
  {Burgess}, \citenamefont {Glorot}, \citenamefont {Botvinick}, \citenamefont
  {Mohamed},\ and\ \citenamefont {Lerchner}}]{higgins_beta-vae_2016}%
  \BibitemOpen
  \bibfield  {author} {\bibinfo {author} {\bibfnamefont {I.}~\bibnamefont
  {Higgins}}, \bibinfo {author} {\bibfnamefont {L.}~\bibnamefont {Matthey}},
  \bibinfo {author} {\bibfnamefont {A.}~\bibnamefont {Pal}}, \bibinfo {author}
  {\bibfnamefont {C.}~\bibnamefont {Burgess}}, \bibinfo {author} {\bibfnamefont
  {X.}~\bibnamefont {Glorot}}, \bibinfo {author} {\bibfnamefont
  {M.}~\bibnamefont {Botvinick}}, \bibinfo {author} {\bibfnamefont
  {S.}~\bibnamefont {Mohamed}}, \ and\ \bibinfo {author} {\bibfnamefont
  {A.}~\bibnamefont {Lerchner}},\ }in\ \href
  {https://openreview.net/forum?id=Sy2fzU9gl} {{\selectlanguage {en}\emph
  {\bibinfo {booktitle} {Proceedings of the 5th International Conference on
  Learning Representations}}}}\ (\bibinfo {address} {Toulon, FR},\ \bibinfo
  {year} {2016})\BibitemShut {NoStop}%
\bibitem [{\citenamefont {Burgess}\ \emph {et~al.}(2017)\citenamefont
  {Burgess}, \citenamefont {Higgins}, \citenamefont {Pal}, \citenamefont
  {Matthey}, \citenamefont {Watters}, \citenamefont {Desjardins},\ and\
  \citenamefont {Lerchner}}]{burgess_understanding_2017}%
  \BibitemOpen
  \bibfield  {author} {\bibinfo {author} {\bibfnamefont {C.~P.}\ \bibnamefont
  {Burgess}}, \bibinfo {author} {\bibfnamefont {I.}~\bibnamefont {Higgins}},
  \bibinfo {author} {\bibfnamefont {A.}~\bibnamefont {Pal}}, \bibinfo {author}
  {\bibfnamefont {L.}~\bibnamefont {Matthey}}, \bibinfo {author} {\bibfnamefont
  {N.}~\bibnamefont {Watters}}, \bibinfo {author} {\bibfnamefont
  {G.}~\bibnamefont {Desjardins}}, \ and\ \bibinfo {author} {\bibfnamefont
  {A.}~\bibnamefont {Lerchner}},\ }in\ \href {http://arxiv.org/abs/1804.03599}
  {\emph {\bibinfo {booktitle} {Proceedings of the 31st Conference on Neural
  Information Processing Systems}}}\ (\bibinfo {address} {Long Beach, USA},\
  \bibinfo {year} {2017})\BibitemShut {NoStop}%
\bibitem [{\citenamefont {Doersch}(2021)}]{doersch_tutorial_2021}%
  \BibitemOpen
  \bibfield  {author} {\bibinfo {author} {\bibfnamefont {C.}~\bibnamefont
  {Doersch}},\ }\href {http://arxiv.org/abs/1606.05908} {\bibfield  {journal}
  {\bibinfo  {journal} {arXiv:1606.05908 [cs, stat]}\ } (\bibinfo {year}
  {2021})}\BibitemShut {NoStop}%
\bibitem [{\citenamefont {Behrisch}\ and\ \citenamefont
  {Eckstein}(2007)}]{behrisch_sputtering_2007}%
  \BibitemOpen
  \bibfield  {author} {\bibinfo {author} {\bibfnamefont {R.}~\bibnamefont
  {Behrisch}}\ and\ \bibinfo {author} {\bibfnamefont {W.}~\bibnamefont
  {Eckstein}},\ }\href {https://dx.doi.org/10.1007/978-3-540-44502-9} {\emph
  {\bibinfo {title} {Sputtering by {Particle} {Bombardment}}}},\ \bibinfo
  {series} {Topics in {Applied} {Physics}}, Vol.\ \bibinfo {volume} {110}\
  (\bibinfo  {publisher} {Springer},\ \bibinfo {address} {Berlin, Germany},\
  \bibinfo {year} {2007})\BibitemShut {NoStop}%
\bibitem [{\citenamefont {Hofsäss}\ \emph {et~al.}(2014)\citenamefont
  {Hofsäss}, \citenamefont {Zhang},\ and\ \citenamefont
  {Mutzke}}]{hofsass_simulation_2014}%
  \BibitemOpen
  \bibfield  {author} {\bibinfo {author} {\bibfnamefont {H.}~\bibnamefont
  {Hofsäss}}, \bibinfo {author} {\bibfnamefont {K.}~\bibnamefont {Zhang}}, \
  and\ \bibinfo {author} {\bibfnamefont {A.}~\bibnamefont {Mutzke}},\ }\href
  {\doibase 10.1016/j.apsusc.2014.03.152} {\bibfield  {journal} {\bibinfo
  {journal} {Applied Surface Science}\ }\textbf {\bibinfo {volume} {310}},\
  \bibinfo {pages} {134} (\bibinfo {year} {2014})}\BibitemShut {NoStop}%
\bibitem [{\citenamefont {Kingma}\ and\ \citenamefont
  {Ba}(2015)}]{kingma_adam_2015}%
  \BibitemOpen
  \bibfield  {author} {\bibinfo {author} {\bibfnamefont {D.~P.}\ \bibnamefont
  {Kingma}}\ and\ \bibinfo {author} {\bibfnamefont {J.}~\bibnamefont {Ba}},\
  }in\ \href {http://arxiv.org/abs/1412.6980} {\emph {\bibinfo {booktitle}
  {Proceedings of the 3rd {International} {Conference} on {Learning}
  {Representations}}}}\ (\bibinfo {address} {San Diego, USA},\ \bibinfo {year}
  {2015})\BibitemShut {NoStop}%
\bibitem [{\citenamefont {Hinton}\ and\ \citenamefont
  {Salakhutdinov}(2006)}]{hinton_reducing_2006}%
  \BibitemOpen
  \bibfield  {author} {\bibinfo {author} {\bibfnamefont {G.~E.}\ \bibnamefont
  {Hinton}}\ and\ \bibinfo {author} {\bibfnamefont {R.~R.}\ \bibnamefont
  {Salakhutdinov}},\ }\href {\doibase 10.1126/science.1127647} {\bibfield
  {journal} {\bibinfo  {journal} {Science}\ }\textbf {\bibinfo {volume}
  {313}},\ \bibinfo {pages} {504} (\bibinfo {year} {2006})}\BibitemShut
  {NoStop}%
\bibitem [{\citenamefont {Abadi}\ \emph {et~al.}(2016)\citenamefont {Abadi},
  \citenamefont {Barham}, \citenamefont {Chen}, \citenamefont {Chen},
  \citenamefont {Davis}, \citenamefont {Dean}, \citenamefont {Devin},
  \citenamefont {Ghemawat}, \citenamefont {Irving}, \citenamefont {Isard},
  \citenamefont {Kudlur}, \citenamefont {Levenberg}, \citenamefont {Monga},
  \citenamefont {Moore}, \citenamefont {Murray}, \citenamefont {Steiner},
  \citenamefont {Tucker}, \citenamefont {Vasudevan}, \citenamefont {Warden},
  \citenamefont {Wicke}, \citenamefont {Yu},\ and\ \citenamefont
  {Zheng}}]{abadi_tensorflow:_2016}%
  \BibitemOpen
  \bibfield  {author} {\bibinfo {author} {\bibfnamefont {M.}~\bibnamefont
  {Abadi}}, \bibinfo {author} {\bibfnamefont {P.}~\bibnamefont {Barham}},
  \bibinfo {author} {\bibfnamefont {J.}~\bibnamefont {Chen}}, \bibinfo {author}
  {\bibfnamefont {Z.}~\bibnamefont {Chen}}, \bibinfo {author} {\bibfnamefont
  {A.}~\bibnamefont {Davis}}, \bibinfo {author} {\bibfnamefont
  {J.}~\bibnamefont {Dean}}, \bibinfo {author} {\bibfnamefont {M.}~\bibnamefont
  {Devin}}, \bibinfo {author} {\bibfnamefont {S.}~\bibnamefont {Ghemawat}},
  \bibinfo {author} {\bibfnamefont {G.}~\bibnamefont {Irving}}, \bibinfo
  {author} {\bibfnamefont {M.}~\bibnamefont {Isard}}, \bibinfo {author}
  {\bibfnamefont {M.}~\bibnamefont {Kudlur}}, \bibinfo {author} {\bibfnamefont
  {J.}~\bibnamefont {Levenberg}}, \bibinfo {author} {\bibfnamefont
  {R.}~\bibnamefont {Monga}}, \bibinfo {author} {\bibfnamefont
  {S.}~\bibnamefont {Moore}}, \bibinfo {author} {\bibfnamefont {D.~G.}\
  \bibnamefont {Murray}}, \bibinfo {author} {\bibfnamefont {B.}~\bibnamefont
  {Steiner}}, \bibinfo {author} {\bibfnamefont {P.}~\bibnamefont {Tucker}},
  \bibinfo {author} {\bibfnamefont {V.}~\bibnamefont {Vasudevan}}, \bibinfo
  {author} {\bibfnamefont {P.}~\bibnamefont {Warden}}, \bibinfo {author}
  {\bibfnamefont {M.}~\bibnamefont {Wicke}}, \bibinfo {author} {\bibfnamefont
  {Y.}~\bibnamefont {Yu}}, \ and\ \bibinfo {author} {\bibfnamefont
  {X.}~\bibnamefont {Zheng}},\ }\href@noop {} {\enquote {\bibinfo {title}
  {{TensorFlow}: {An} {Open} {Source} {Machine} {Learning} {Framework} for
  {Everyone}},}\ } (\bibinfo {year} {2016}),\ \bibinfo {note}
  {https://tensorflow.org/}\BibitemShut {NoStop}%
\bibitem [{\citenamefont {Chollet}\ and\ \citenamefont
  {{others}}(2015)}]{chollet_keras:_2015}%
  \BibitemOpen
  \bibfield  {author} {\bibinfo {author} {\bibfnamefont {F.}~\bibnamefont
  {Chollet}}\ and\ \bibinfo {author} {\bibnamefont {{others}}},\ }\href@noop {}
  {\enquote {\bibinfo {title} {Keras: {The} {Python} {Deep} {Learning}
  library},}\ } (\bibinfo {year} {2015}),\ \bibinfo {note}
  {https://keras.io/}\BibitemShut {NoStop}%
\bibitem [{\citenamefont {Sohn}\ \emph {et~al.}(2015)\citenamefont {Sohn},
  \citenamefont {Lee},\ and\ \citenamefont {Yan}}]{sohn_learning_2015}%
  \BibitemOpen
  \bibfield  {author} {\bibinfo {author} {\bibfnamefont {K.}~\bibnamefont
  {Sohn}}, \bibinfo {author} {\bibfnamefont {H.}~\bibnamefont {Lee}}, \ and\
  \bibinfo {author} {\bibfnamefont {X.}~\bibnamefont {Yan}},\ }in\ \href
  {https://papers.nips.cc/paper/2015/hash/8d55a249e6baa5c06772297520da2051-Abstract.html}
  {\emph {\bibinfo {booktitle} {Proceedings of the 29th Conference on Neural
  Information Processing Systems}}},\ Vol.~\bibinfo {volume} {28}\ (\bibinfo
  {address} {Montréal, CA},\ \bibinfo {year} {2015})\BibitemShut {NoStop}%
\end{thebibliography}%

\end{document}